\documentclass[10pt, final, journal, letterpaper, twocolumn]{IEEEtran}
\makeatletter
\def\ps@headings{%
\def\@oddhead{\mbox{}\scriptsize\rightmark \hfil \thepage}%
\def\@evenhead{\scriptsize\thepage \hfil \leftmark\mbox{}}%
\def\@oddfoot{}%
\def\@evenfoot{}}
\makeatother 

\IEEEoverridecommandlockouts

\usepackage{amsfonts}
\usepackage[dvips]{graphicx}
\usepackage{caption}
\usepackage{subfigure}
\usepackage{times}
\usepackage{cite}
\usepackage{lettrine}
\usepackage{amsmath}
\usepackage{amsmath}
\usepackage{amsthm}
\allowdisplaybreaks[4]
\usepackage{array}
\usepackage{amssymb}

\usepackage{stfloats}
\usepackage{slashbox}
\usepackage{graphicx}
\usepackage{footnote}
\usepackage{booktabs}
\usepackage{array}
\usepackage{algorithmic}
\usepackage{algorithm}
\usepackage{subeqnarray}
\usepackage{cases}
\usepackage{threeparttable}
\usepackage{color}

\begin{document}

\title{Optimal Task Offloading and Resource Allocation in Mobile-Edge Computing with Inter-user Task Dependency}

\IEEEoverridecommandlockouts

\author{Jia~Yan,~\IEEEmembership{Student Member,~IEEE}, Suzhi~Bi,~\IEEEmembership{Senior Member,~IEEE},  Ying-Jun~Angela~Zhang,~\IEEEmembership{Senior Member,~IEEE}, and  Meixia Tao,~\IEEEmembership{Fellow,~IEEE}
\thanks{This work has been presented in part at the IEEE Global Communications Conference (GLOBECOM), Abu Dhabi, UAE, Dec. 9-13, 2018 \cite{myglobecom}. 

J. Yan (yj117@ie.cuhk.edu.hk) and Y. J. Zhang (yjzhang@ie.cuhk.edu.hk) are with the Department of Information Engineering, The Chinese University of Hong Kong, Hong Kong.

S. Bi (bsz@szu.edu.cn) is with the College of Electronic and Information Engineering, Shenzhen University, Shenzhen, China.

M. Tao (mxtao@ieee.org) is with the Department of Electronic Engineering, Shanghai Jiao Tong University, Shanghai, China.}
}

\maketitle

\vspace{-1.5cm}

\begin{abstract}
   Mobile-edge computing (MEC) has recently emerged as a cost-effective paradigm to enhance the computing capability of hardware-constrained wireless devices (WDs). In this paper, we first consider a two-user MEC network, where each WD has a sequence of tasks to execute. In particular, we consider task dependency between the two WDs, where the input of a task at one WD requires the final task output at the other WD. Under the considered task-dependency model, we study the optimal task offloading policy and resource allocation (e.g., on offloading transmit power and local CPU frequencies) that minimize the weighted sum of the WDs' energy consumption and task execution time. The problem is challenging due to the combinatorial nature of the offloading decisions among all tasks and the strong coupling with resource allocation. To tackle this problem, we first assume that the offloading decisions are given and derive the closed-form expressions of the optimal offloading transmit power and local CPU frequencies. Then, an efficient bi-section search method is proposed to obtain the optimal solutions. Furthermore, we prove that the optimal offloading decisions follow an one-climb policy, based on which a reduced-complexity Gibbs Sampling algorithm is proposed to obtain the optimal offloading decisions. We then extend the investigation to a general multi-user scenario, where the input of a task at one WD requires the final task outputs from multiple other WDs. Numerical results show that the proposed method can significantly outperform the other representative benchmarks and efficiently achieve low complexity with respect to the call graph size.

\end{abstract}

\section{Introduction}
The explosive growth of Internet of Things (IoT) in recent years enables cost-effective interconnections between tens of billions of wireless devices (WDs), such as sensors and wearable devices. Due to the stringent size constraint and production cost concern, an IoT device is often equipped with a limited battery and a low-performance on-chip computing unit, which are recognized as two fundamental impediments for supporting computation intensive applications in future IoT. Mobile edge computing (MEC) \cite{MECsurvey1,MECsurvey2}, viewed as an efficient solution, has attracted significant attention. The key idea of MEC is to offload intensive computation tasks to the edges of radio access network, where much more powerful servers will compute on behalf of the resource-limited WDs. Compared with the traditional mobile cloud computing, MEC can overcome the drawbacks of high overhead and long backhaul latency.

In general, MEC has two computation offloading models: binary and partial offloading \cite{MECsurvey1}. Binary offloading requires each task to be either computed locally or offloaded to the MEC server as a whole. Partial offloading, on the other hand, allows a task to be partitioned and executed both locally and at the MEC server. In this paper, we consider binary computation offloading, which is commonly used in IoT systems for processing non-partitionable simple tasks \cite{bi_DRL,MEC3}.

Due to the time-varying wireless channel fading, it is not necessarily optimal to always offload all the computations to the MEC server, e.g., deep fading may lead to very low offloading data rate. Meanwhile, wireless resource allocation, e.g., transmit time and power, needs to be jointly designed with the computation offloading for optimum computing performance. In this regard, on the one hand, \cite{MEC3,xu,MEC2,MEC5} focused on the optimal binary offloading policies when each user only has one task to be executed. Specifically, \cite{MEC5} considered energy-optimal offloading and resource allocation in the single user case. Authors in \cite{MEC2} further considered a wireless powered MEC and maximize the probability of successful computations. The performance optimization of multi-user wireless powered MEC system was later studied in \cite{MEC3,xu}. On the other hand, \cite{MEC7,MEC4,HongXing} considered a more general scenario, where the binary offloading model is applied to multiple independent tasks. Specifically, \cite{MEC7} considered multiple mobile users that each offloads multiple independent tasks to a single access point. In \cite{MEC4}, a single user can offload independent tasks to multiple edge devices, which then minimizes the weight sum of WD's energy consumption and total tasks' execution latency. In \cite{HongXing}, the user offloads independent tasks to the edge devices and downloads results from them over pre-scheduled time slots. Energy consumption at both the user and edge devices is considered therein.

Nonetheless, the above studies do not consider the important dependency among different tasks in various applications. That is, a user often needs to execute multiple related tasks, where the input of one task requires the output of another. Since the executions are coupled among tasks, the optimal design becomes much more difficult than the previous case where independent tasks can be executed in parallel. Call graphs \cite{cs_taskgraph} are commonly used to model the dependency among different tasks \cite{add1,add2,add3,single1,single4,single5,single2,multi1}. \cite{add1,add2,add3} considered the cloud computing environments with multiple virtual machines (VMs) and aimed to map the tasks in a general call graph to the VMs by minimizing the overall execution cost while meeting deadline constraint. For a single-user MEC system, \cite{single1} considered a general call graph and obtained the joint optimal task-offloading decisions and transmit power that minimize the WD's energy consumption under latency constraint. Besides, the authors in \cite{single4} considered a sequential call graph for a single user and derived an optimal one-climb policy, which means that the execution migrates only at most once between the WD and the cloud server. This work was extended to a call graph with a general topology in \cite{single5} and a heuristic task offloading problem was studied in \cite{single2}. A multi-user case was considered in \cite{multi1}, where each independent WD has multiple tasks with a general call graph and the goal is to optimize the energy efficiency. Notice that the above work \cite{single1,single4,single5,single2,multi1} considered a non-causal channel model that assumes perfect knowledge of time-varying channel conditions throughout the task executions in order to derive the optimal structure of the offloading decisions in their considered task call graphs.

The call graphs considered by most of the existing studies on MEC, such as in \cite{single1,single4,single5,single2,multi1}, only take into account the dependency among tasks executed by an individual WD. In practice, tasks executed by different WDs usually have relevance as well. For example, an IoT sensor often needs to combine
the processed data from other sensors. Consider a smart home environment where a wireless sensor keeps measuring the temperature of the room and processes the sensed raw data through a series of operations. The obtained temperature estimation is useful for controlling other smart home appliances, e.g., air conditioner and aquarium heating device. Meanwhile, a wireless controller has the function of sensing the air humidity and controls the air conditioner (i.e., temperature setting and service hours) according to its own processed air humidity data and the room temperature data estimated by the wireless sensor. Another example is distributed learning and inference in wireless sensor networks. For parametric estimation problems, \cite{DL2} considered one sensor passing its quantized estimation result to the other sensor, which subsequently generates its own estimation result by jointly processing its local observation and the received quantized data. For nonparametric approaches, in \cite{DL}, a sensor can use the computation results shared by the other sensors to compute a global estimation for least-squares regression through massage-passing algorithms. The inter-user task dependency has significant impact to the offloading and resource allocation decisions. For instance, a WD is likely to offload its task to the edge server even when the channel condition is poor, because another WD with time-critical applications is urgently in need of its computation output. Besides, the exchange of computation results for dependent tasks also consumes extra energy and time. In general, the case with inter-user dependency requires the joint optimization of tasks executions of all correlated users, which is a challenging problem yet lacking of concrete study.

\begin{figure*}[t]
\begin{centering}
\includegraphics[scale=0.6]{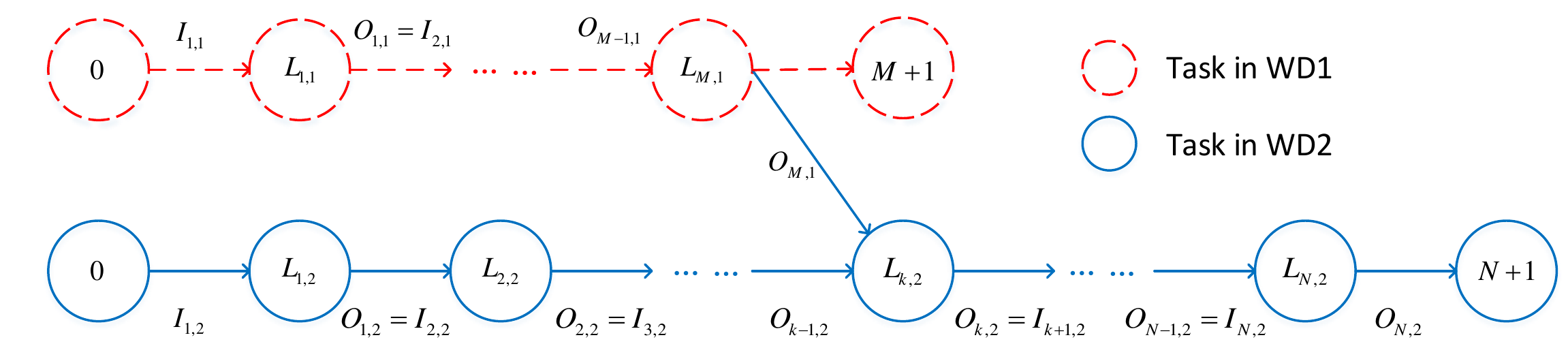}
\vspace{-0.1cm}
 \caption{The considered call graph in a two-user MEC network. }\label{fig1}
\end{centering}
\vspace{-0.1cm}
\end{figure*}

In this paper, we consider a task call graph in a two-user MEC system as shown in Fig. \ref{fig1}, where the computation of an intermediate task at WD2 requires the output of the last task at WD1. To the authors' best knowledge, this is the first work that exploits the task dependency across different users in an MEC system. As a first step to study the inter-user task dependency in an MEC system, we will first consider a simplified two-user model to capture the optimal solution properties and the impact of user dependency to the optimal system performance. Then, we will extend the investigation to a multi-user scenario as shown in Fig. 5. The main contributions of this paper are as follows:
\begin{itemize}
  \item  With the inter-user task dependency in Fig. \ref{fig1}, we formulate a mixed integer optimization problem to minimize the weighted sum of the WDs' energy consumption and task execution time. The task offloading decisions, local CPU frequencies and transmit power of each WD are jointly optimized. The problem is challenging due to the combinatorial nature of the offloading decisions among all tasks in such call graph and the strong coupling with resource allocation.
\item Given the offloading decisions, we first derive closed-form solutions of the optimal local CPU frequencies and transmit power of each WD, respectively. We then establish an inequality condition of the completion time between the two dependent tasks, based on which an efficient bi-section search method is proposed to obtain the optimal resource allocation.
\item We show that the optimal offloading decisions follow an one-climb policy, where each WD offloads its data at most once to the edge server at the optimum. Based on the one-climb policy, we propose a reduced-complexity Gibbs sampling algorithm to obtain the optimal offloading decisions.

\item We further extend the study to a general multi-user scenario, where the input of a task at one WD requires the final task outputs from multiple other WDs. Then, we prove that the one-climb policy is still optimal for each user. Besides, the one-climb based Gibbs sampling algorithm is adapted to solve the offloading decision optimization problem in the multi-user scenario.

\end{itemize}

Simulation results show that our proposed algorithm can effectively reduce the energy consumption and computation delay compared with other representative benchmarks. In particular, it significantly outperforms the scheme that neglects the task dependency and optimizes the two WDs' performance individually. Meanwhile, the proposed method has low computational complexity with respect to the size of call graph. It is worth mentioning that this paper assumes non-causal channel information, where the AP is assumed to have full channel state information (CSI) when uploading/downloading the tasks. The assumption allows us to analyze the properties of the optimal solution, especially the structure of the optimal offloading decisions under inter-user dependency. These properties are useful in the future design of online algorithms that consider more practical channel prior knowledge. Meanwhile, the proposed solution method provides an offline performance benchmark for evaluating online offloading strategies that will be investigated in the future.

The rest of the paper is organized as follows. In Section II, we describe the system model and formulate the problem. The optimal CPU frequencies and transmit power of each WD under fixed offloading decisions are derived in Section III. In Section IV, we first prove that the optimal offloading decisions follow an one-climb policy and based on that, a reduced-complexity Gibbs sampling algorithm is proposed. We extend the study to a general multi-user case in Section V.  In Section VI, the performance of the proposed algorithms is evaluated via simulations. Finally, we conclude the paper in Section VII.


\section{System Model and Problem Formulation}

We first consider an MEC system with two WDs and one access point (AP), all equipped with single antenna. We will extend to a system with multiple WDs in Section V. The AP is the gateway of the edge cloud and has a stable power supply. As shown in Fig.1, WD1 and WD2 have $M$ and $N$ sequential tasks to execute, respectively. For simplicity of exposition, we introduce for each WD an auxiliary node 0 as the entry task, and auxiliary nodes $M+1$, $N+1$ as the exit tasks for WD1 and WD2, respectively. In particular, we assume that the computations of the two WDs are related, such that the calculation of an intermediate task of WD2, denoted as $k$, for $k=1,...,N$, requires the output of the last task $M$ of WD1.

Each task $i$ of WD $j$ is characterized by a three-item tuple $(L_{i,j},I_{i,j},O_{i,j})$, where $i=0,1,...,M+1$ when $j=1$, and $i=0,1,...,N+1$ when $j=2$. Specifically, $L_{i,j}$ denotes the computing workload in terms of the total number of CPU cycles required for accomplishing the task, $I_{i,j}$ and $O_{i,j}$ denote the size of computation input and output data in bits, respectively. As for the two auxiliary nodes of each WD, $L_{0,j}=L_{M+1,1}=L_{N+1,2}=0$. For WD1, it holds that $I_{i,1}=O_{i-1,1}$, $i=1,...,M+1$. As for the WD2, we have
\begin{align}
I_{i,2}=\left\{
          \begin{array}{ll}
            O_{i-1,2}+O_{M,1}, & i=k, \\
            O_{i-1,2}, & \hbox{otherwise.}
          \end{array}
        \right.
\end{align}
Moreover, $I_{i,j}=0$ for the entry node and $O_{i,j}=0$ for the exit node of each WD.

We assume that the two series of tasks must be initiated and terminated at the respective WD. That is, the auxiliary entry and exist tasks must be executed locally, while the other $(M+N)$ actual tasks can be either executed locally or offloaded to the edge server. We denote the computation offloading decision of task $i$ of WD $j$ as $a_{i,j}\in\{0,1\}$, where $a_{i,j}=1$ denotes edge execution and $a_{i,j}=0$ denotes local computation.

In addition, we assume that each WD is allocated with an orthogonal channel of equal bandwidth $W$, thus there is no interference between the WDs when offloading/downloading. The wireless channel gains between the WD $j$ and the AP when offloading and downloading task $i$ are denoted as $h_{i,j}$ and $g_{i,j}$, respectively. Besides, we assume additive white Gaussian noise (AWGN) with zero mean and equal variance $\sigma^{2}$ at all receivers for each user.

\emph{\textbf{Remark 1:}} In many low-power IoT systems, e.g., wireless sensor networks, the data rate for task offloading is not demanding (e.g., tens to several hundred kbps) and the required bandwidth is usually small \cite{sensor1}. For instance, in the narrowband Internet of Things (NB-IoT) system, a 10MHz LTE carrier can supply orthogonal transmissions of more than 50 users \cite{sensor2}. In addition, in some mobile communication systems such as LTE, each user is allocated a dedicated resource block throughout its transmission. Besides, according to the one-climb offloading property derived in Section IV, each WD offloads its data at most once to the edge server at the optimum, which indicates that the chance of the two WDs offloading at the same time to contend for bandwidth is very small in general. \emph{Therefore, in this paper, we assume that each device is allocated with an orthogonal channel of equal bandwidth when the number of WDs is moderate [9,10,20].}


In the following, we discuss the computation overhead in terms of execution time and energy consumption for local and edge computing, respectively.

\subsection{Local Computing}

We denote the CPU frequency of WD $j$ for computing task $i$ as $f_{i,j}$. Thus, the local computation execution time can be given by
\begin{align}\label{f_mapping}
\tau_{i,j}^{l}=\frac{L_{i,j}}{f_{i,j}},
\end{align}
and the corresponding energy consumption is \cite{MECsurvey1}
\begin{align}\label{e_local}
e_{i,j}^{l}=\kappa L_{i,j}f_{i,j}^{2}=\kappa\frac{(L_{i,j})^{3}}{(\tau_{i,j}^{l})^{2}},
\end{align}
where $\kappa$ is the fixed effective switched capacitance parameter depending on the chip architecture.

\subsection{Edge Computing}

Let $p_{i,j}$ denote the transmit power of WD $j$ when offloading task $i$ to the AP, and we can express the uplink data rate for offloading task $i$ of WD $j$ as
\begin{align}\label{rate_up}
R_{i,j}^{u}=W\log_{2}\left(1+\frac{p_{i,j}h_{i,j}}{\sigma^{2}}\right).
\end{align}

From \eqref{rate_up}, the transmission time of WD $j$ when offloading task $i$ is expressed as
\begin{align}\label{t_trans}
\tau_{i,j}^{u}=\frac{O_{i-1,j}}{R_{i,j}^{u}}.
\end{align}
Define $f(x)\triangleq\sigma^{2}\left(2^{\frac{x}{W}}-1\right)$. It follows from \eqref{rate_up} and \eqref{t_trans} that
\begin{align}\label{p_mapping}
p_{i,j}=\frac{1}{h_{i,j}}f\left(\frac{O_{i-1,j}}{\tau_{i,j}^{u}}\right).
\end{align}
Then, the transmission energy consumption is
\begin{align}\label{e_edge}
e_{i,j}^{u}=p_{i,j}\tau_{i,j}^{u}=\frac{\tau_{i,j}^{u}}{h_{i,j}}f\left(\frac{O_{i-1,j}}{\tau_{i,j}^{u}}\right).
\end{align}
Notice that \eqref{e_edge} is convex in $\tau_{i,j}^u$ since \eqref{e_edge} is the perspective function with respect to $\tau_{i,j}^u$ of a convex function $f(x)$ \cite{convex}.

The execution time of task $i$ of WD $j$ on the edge is given by $\tau_{i,j}^{c}=\frac{L_{i,j}}{f_{c}}$, where $f_c$ is the constant CPU frequency of the edge server.

\emph{\textbf{Remark 2:}} Since we consider that each WD has a sequence of tasks to execute in the MEC network, there are at most $J$ tasks being computed at the AP simultaneously, where $J$ is the number of WDs.  Besides, in practice, the server located at the AP is usually multi-core [9-11,16,20], thus can handle the $J$ tasks at the same time when $J$ is moderate.  \emph{In this paper, we assume that the AP has a multi-core processor and each core has a fixed service rate $f_c$ assigned to process one task.}

Furthermore, as for the downlink transmission, we denote the fixed transmit power of the AP by $P_0$. Thus, the downlink data rate for feeding the $i$-th task's input of WD $j$ from the AP when computing task $i$ locally can be expressed as
\begin{align}
R_{i,j}^{d}=W\log_{2}\left(1+\frac{P_{0}g_{i,j}}{\sigma^{2}}\right).
\end{align}

Likewise, the time needed for the downlink transmission is given by $\tau_{i,j}^{d}=\frac{O_{i-1,j}}{R_{i,j}^{d}}$.

\begin{figure}
\begin{centering}
\includegraphics[scale=0.6]{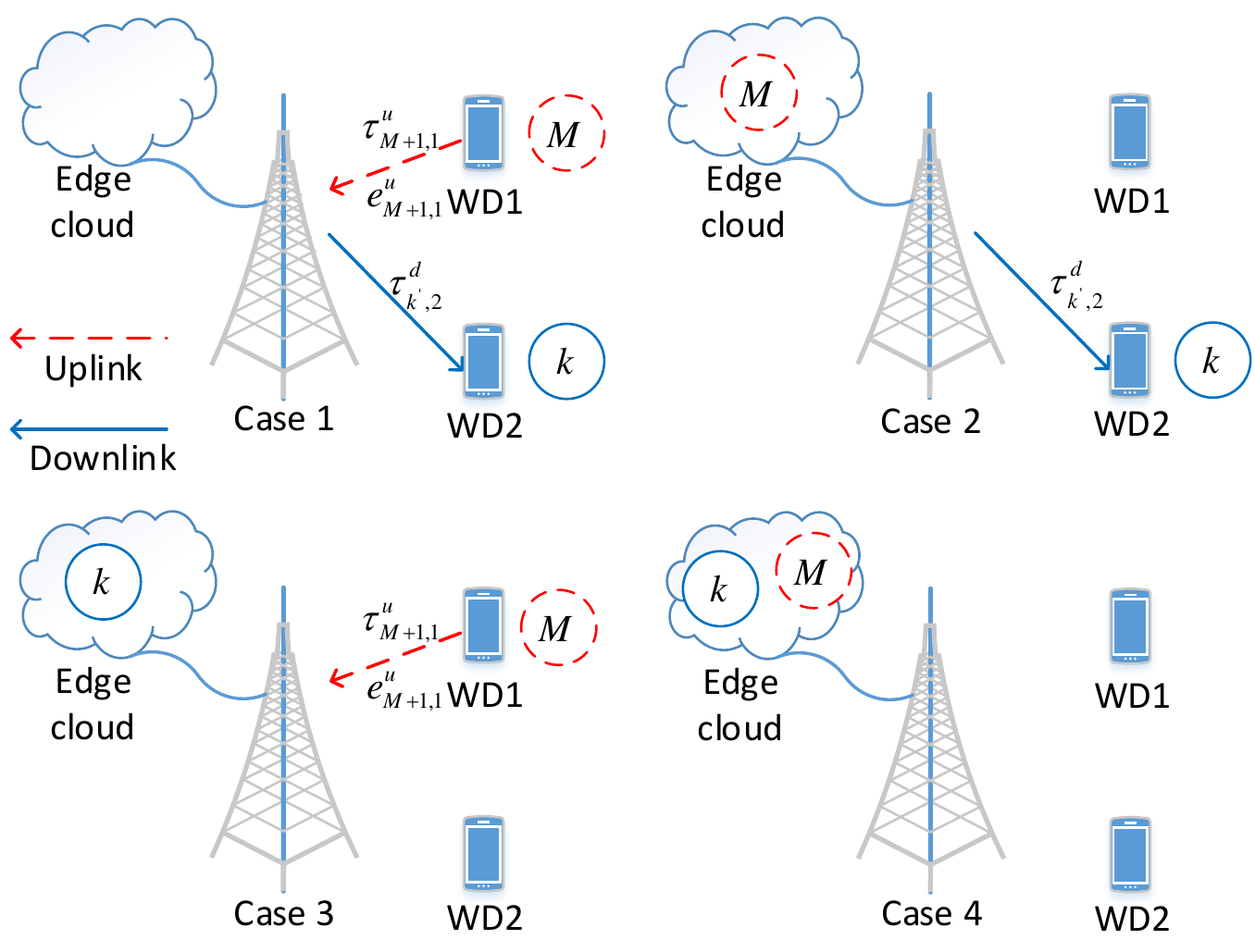}\label{relationship}
\vspace{-0.1cm}
 \caption{Illustration of the task dependency between two WDs. }
\end{centering}
\vspace{-0.1cm}
\end{figure}

\subsection{Task Dependency Model }

As shown in Fig. 2, the task dependency model between the two WDs can be one of the following four cases, depending on the values of $a_{M,1}$ and $a_{k,2}$.

\begin{itemize}
  \item \textbf{Case 1:} When both the $M$-th task of WD1 and the $k$-th task of WD2 are executed locally, i.e., $a_{M,1}=0$ and $a_{k,2}=0$, the AP acts as a relay node. First, the WD1 uploads its output of $M$-th task to the AP. Then, the AP forwards this information to the WD2. Specifically, the uplink transmission time and energy in this process are
      \begin{align}
\tau_{M+1,1}^{u}=\frac{O_{M,1}}{R_{M+1,1}^{u}}
\end{align}
and
      \begin{align}
e_{M+1,1}^{u}=p_{M+1,1}\tau_{M+1,1}^{u},
\end{align}
       respectively, where $R_{M+1,1}^{u}$ and $p_{M+1,1}$ are the corresponding uplink data rate and uplink transmit power, respectively. As for the downlink transmission, the transmission time is denoted as
             \begin{align}
\tau_{k',2}^d=\frac{O_{M,1}}{R_{k,2}^{d}}.
\end{align}
  \item \textbf{Case 2:} When the $M$-th task of WD1 is executed at the edge and the $k$-th task of WD2 is computed locally, i.e., $a_{M,1}=1$ and $a_{k,2}=0$, the output of $M$-th task of WD1 is downloaded to the WD2 after execution at the edge. 
  \item \textbf{Case 3:} In this case, the $M$-th task of WD1 is executed locally and the $k$-th task of WD2 is offloaded to the edge, i.e., $a_{M,1}=0$ and $a_{k,2}=1$. The WD1 needs to upload the result before the computation of the $k$-th task of WD2 at the edge. 
  \item \textbf{Case 4:} In this case, both the $M$-th task of WD1 and the $k$-th task of WD2 are executed at the edge, i.e., $a_{M,1}=1$ and $a_{k,2}=1$. Therefore, neither uplink nor downlink transmission is needed.
\end{itemize}

\subsection{Problem Formulation}
From the above discussion, in order to obtain the total tasks execution time of WD1, we first denote the time spent on computations both locally and at the edge server by $T_1^{comp}$, which can be expressed as
\begin{align}
T_{1}^{comp}=\sum_{i=1}^{M}\left[(1-a_{i,1})\tau_{i,1}^{l}+a_{i,1}\tau_{i,1}^{c}\right].
\end{align}
As for the communication delay $T_1^{tran}$ consumed on offloading/downloading the task data to/from the AP, we have
\begin{align}\nonumber
T_{1}^{tran}&=\sum_{i=1}^{M+1}\left[a_{i,1}(1-a_{i-1,1})\tau_{i,1}^{u}+(1-a_{i,1})a_{i-1,1}\tau_{i,1}^{d}\right]\nonumber\\
&=\sum_{i=1}^{M+1}\left[a_{i,1}\tau_{i,1}^{u}+a_{i-1,1}\tau_{i,1}^{d}-a_{i-1,1}a_{i,1}(\tau_{i,1}^{u}+\tau_{i,1}^{d})\right].
\end{align}
%
Note that there is no communication delay for the $i$-th task if $a_{i-1,1}=a_{i,1}$, i.e., the two tasks are computed at the same device. Otherwise, if $a_{i-1,1}=0$ and $a_{i,1}=1$, the communication delay is due to the uplink transmission time $\tau_{i,1}^{u}$, whereas, if $a_{i-1,1}=1$ and $a_{i,1}=0$, the communication delay is due to the downlink transmission time $\tau_{i,1}^{d}$. Therefore, the total tasks execution time of WD1 is
\begin{align}
T_{1}=T_{1}^{comp}+T_{1}^{tran}.
\end{align}

Furthermore, we can calculate the total energy consumption of WD1 by
\begin{align}\label{e1}
E_{1}=&\sum_{i=1}^{M}\left[(1-a_{i,1})e_{i,1}^{l}+a_{i,1}(1-a_{i-1,1})e_{i,1}^{u}\right]\nonumber\\
&+(1-a_{M,1})e_{M+1,1}^{u},
\end{align}
which consists of the total execution energy of $M$ tasks and the energy consumption on offloading the final result if the $M$-th task is computed locally, i.e., when $a_{M,1}=0$. Note that the energy cost for the uplink transmission $e_{i,1}^{u}, i\in\{1,...,M\}$ occurs in \eqref{e1} only if $a_{i,1}=1$ and $a_{i-1,1}=0$.

Similarly, the total computation energy consumption of WD2 can be expressed as
\begin{align}\label{e2}
E_{2}=\sum_{i=1}^{N}\left[(1-a_{i,2})e_{i,2}^{l}+a_{i,2}(1-a_{i-1,2})e_{i,2}^{u}\right].
\end{align}
As for the execution time of WD2, we first consider the waiting time until the output of the $M$-th task of WD1 reaches WD2, denoted by $T_1^{wait}$, as follows.
\begin{align}\label{WT1}
T^{wait}_{1}=&\sum_{i=1}^{M}\bigg[(1-a_{i,1})\tau_{i,1}^{l}+a_{i,1}(\tau_{i,1}^{c}+\tau_{i,1}^{u})\nonumber\\
&+a_{i-1,1}\tau_{i,1}^{d}-a_{i-1,1}a_{i,1}(\tau_{i,1}^{u}+\tau_{i,1}^{d})\bigg]
\nonumber\\&+(1-a_{M,1})\tau_{M+1,1}^{u}+(1-a_{k,2})\tau_{k',2}^{d}.
\end{align}
It consists of the total execution time of $M$ tasks of WD1, and the transmit time of the output of the $M$-th task as shown in the four cases of Fig. 2.

Meanwhile, the waiting time until the output of the $(k-1)$-th task of WD2 is ready, denoted by $T_2^{wait}$, is given by
\begin{align}\label{WT2}
T^{wait}_{2}=&\sum_{i=1}^{k-1}\bigg[(1-a_{i,2})\tau_{i,2}^{l}+a_{i,2}(\tau_{i,2}^{c}+\tau_{i,2}^{u})\nonumber\\
&+a_{i-1,2}\tau_{i,2}^{d}-a_{i-1,2}a_{i,2}(\tau_{i,2}^{u}+\tau_{i,2}^{d})\bigg]
\nonumber\\&+a_{k,2}\tau_{k,2}^{u}+a_{k-1,2}\tau_{k,2}^{d}-a_{k-1,2}a_{k,2}(\tau_{k,2}^{u}+\tau_{k,2}^{d}),
\end{align}
which includes the total execution time of the first $k-1$ tasks and the transmission time on offloading task $k$ (i.e., when $a_{k-1,2}=0$, $a_{k,2}=1$) or downloading the output of task $(k-1)$ to WD2 (i.e., when $a_{k-1,2}=1$, $a_{k,2}=0$). From \eqref{WT1} and \eqref{WT2}, the total waiting time before the $k$-th task of WD2 is ready for execution is
\begin{align}
T^{wait}=\max\left\{T^{wait}_{1},T^{wait}_{2}\right\}.
\end{align}
Accordingly, the total task execution time of WD2 equals to $T^{wait}$ plus the execution time of tasks from $k$ to $N$, i.e.,
\begin{align}\label{T2}
T_{2}=&T^{wait}+\sum_{i=k}^{N}\left[(1-a_{i,2})\tau_{i,2}^{l}+a_{i,2}\tau_{i,2}^{c}\right]\nonumber\\
&+\sum_{i=k+1}^{N+1}\bigg[a_{i,2}\tau_{i,2}^{u}+a_{i-1,2}\tau_{i,2}^{d}-a_{i-1,2}a_{i,2}(\tau_{i,2}^{u}+\tau_{i,2}^{d})\bigg].
\end{align}

In this paper, we consider the energy-time cost (ETC) as the performance metric \cite{MECsurvey1,MEC4}, which is defined as the weighted sum of total energy consumption and execution time, i.e.,
\begin{align}
\eta_{1}=\beta_{1}^{E}E_{1}+\beta_{1}^{T}T_{1},
\end{align}
where $0<\beta_{1}^{E}<1$ and $0\leq\beta_{1}^{T}<1$ denote the weights of energy consumption and computation completion time for WD1, respectively. Without loss of generality, it is assumed that the weights are related by $\beta_{1}^{E}=1-\beta_{1}^{T}$.
Accordingly, the ETC of WD2 is
\begin{align}
\eta_{2}=\beta_{2}^{E}E_{2}+\beta_{2}^{T}T_{2},
\end{align}
where $0<\beta_{2}^{E}<1$ and $0<\beta_{2}^{T}<1$ denote the two weighting parameters satisfying $\beta_{2}^{E}=1-\beta_{2}^{T}$. It is worth noting that $\beta_{1}^{T}=0$ represents a special case which will be discussed in Section III, while $\beta_{2}^{T}=0$ leads to a trivial solution that the WD2 will take infinitely long time to finish its task executions. In practice, we allow different WDs to choose different wights to meet user-specific demands. For example, a WD $j$ with delay-sensitive applications, such as watching movies and online game, prefers to choose a larger $\beta_{j}^{T}$ to reduce the delay. Besides, a WD $j$ with low battery energy tends to set a larger $\beta_{j}^{E}$ to save more energy.

Denoting $\mathbf{a}\triangleq\{a_{i,j}\}$, $\mathbf{p}\triangleq\{p_{i,j}\}$, and $\mathbf{f}\triangleq\{f_{i,j}\}$, we are interested in minimizing the total ETC of the two WDs by solving the following problem:
\begin{eqnarray}
\mbox{(P1)}~~\min_{(\mathbf{a},\mathbf{p},\mathbf{f})}&&\eta_{1}+\eta_{2},\nonumber \\
{\rm s.t.}&&0\leq p_{i,j}\leq P_{peak},\nonumber \\
&&0\leq f_{i,j}\leq f_{peak},\nonumber \\
&&a_{i,j}\in\{0,1\},\forall i,j,
\end{eqnarray}
where the first two constraints correspond to the peak transmit power and peak CPU frequency. We assume $f_c>f_{peak}$ in this paper. For practical implementation, it is assumed that there exists a controller at the AP to obtain the optimal offloading and resource allocation decisions by solving (P1). Then, each WD can receive the control signal from the controller at the AP to perform optimal task offloading and resource allocation. Because of the one-to-one mappings between $f_{i,j}$ and $\tau_{i,j}^l$ in \eqref{f_mapping} and between $p_{i,j}$ and $\tau_{i,j}^u$ in \eqref{p_mapping}, it is equivalent to optimize (P1) over the time allocation $(\tau_{i,j}^l,\tau_{i,j}^u)$.  By introducing an auxiliary variable $t=\max\left\{T^{wait}_{1},T^{wait}_{2}\right\}$, (P1) can be equivalently expressed as
\begin{eqnarray}
\mbox{(P2)}~~\min_{(\mathbf{a},\{\tau_{i,j}^{u}\},\{\tau_{i,j}^{l}\},t)}&&\eta_{1}+\eta_{2},\nonumber\\
{\rm s.t.}&&t\geq T^{wait}_{1},t\geq T^{wait}_2,\nonumber\\
&&\tau_{i,j}^{u}\geq\frac{O_{i-1,j}}{W\log_{2}\left(1+\frac{P_{peak}h_{i,j}}{\sigma^{2}}\right)},\nonumber\\
&&\tau_{i,j}^l\geq\frac{L_{i,j}}{f_{peak}},\nonumber\\
&&a_{i,j}\in\{0,1\},\forall i,j.
\end{eqnarray}

Suppose that we have obtained the optimal solution $\{\mathbf{a}^*,(\tau_{i,j}^{u})^*,(\tau_{i,j}^{l})^*\}$ of (P2). Then, we can easily retrieve the unique $f^*_{i,j}$ and $p^*_{i,j}$ in (P1) using \eqref{f_mapping} and \eqref{p_mapping}, respectively. Notice that (P2) is non-convex in general due to the binary variables $\mathbf{a}$. However, it can be seen that for any given $\mathbf{a}$, the remaining optimization over $(\tau_{i,j}^{l},\tau_{i,j}^{u},t)$ is a convex problem. In the following section, we assume that the offloading decision $\mathbf{a}$ is given and study some interesting properties of the optimal CPU frequencies and the transmit power of each WD, based on which an efficient method is proposed to obtain the optimal solutions.

\section{Optimal Resource Allocation under Fixed Offloading Decision}

\subsection{Optimal Solution of (P2) given $\mathbf{a}$}
Suppose that $\mathbf{a}$ is given. A partial Lagrangian of Problem (P2) is given by
\begin{align}\label{lagrangian}
L(\{\tau_{i,j}^{u}\},\{\tau_{i,j}^{l}\},t,\lambda,\mu)=&\eta_{1}+\eta_{2}+\lambda\left(T^{wait}_1-t\right)\nonumber\\
&+\mu\left(T^{wait}_2-t\right),
\end{align}
where $\lambda\geq 0$ and $\mu\geq 0$ denote the dual variables associated with the corresponding constraints.

Let $\lambda^{*}$ and $\mu^{*}$ denote the optimal dual variables. We derive the closed-form expressions of the optimal CPU frequencies and transmit power of each WD as follows.

\emph{\textbf{Proposition 3.1:}} $\forall i,j$ with $a_{i,j}=0$, the optimal CPU frequencies of the two WDs satisfy
\begin{align}\label{f1}
f_{i,1}^{*}=\min\left\{\sqrt[3]{\frac{\beta_{1}^{T}+\lambda^{*}}{2\kappa\beta_{1}^{E}}},f_{peak}\right\},\forall i\in\{1,..,M\},
\end{align}
\begin{align}\label{f2}
f_{i,2}^{*}=\left\{
              \begin{array}{ll}
                \min\left\{\sqrt[3]{\frac{\mu^{*}}{2\kappa\beta_{2}^{E}}},f_{peak}\right\}, & i\in\{1,...,k-1\},
 \\
               \min\left\{\sqrt[3]{\frac{\beta_{2}^{T}}{2\kappa\beta_{2}^{E}}},f_{peak}\right\}, & i\in\{k,...,N\} .
              \end{array}
            \right.
\end{align}

\begin{proof}
Please refer to Appendix \ref{appendicesA}.
\end{proof}

From Proposition 3.1, we have the following observations:
\begin{itemize}
  \item The optimal local CPU frequencies are the same for all the tasks of the same type, i.e., $i\in\{1,...,M\}$ in WD1, $i\in\{1,...,k-1\}$ or $i\in\{k,...,N\}$ in WD2, regardless of the wireless channel conditions and workloads.
  \item For each task of WD1, when $\beta_1^T$ or $\lambda^*$ increases (a larger $\lambda^*$ corresponds to a tighter task dependency constraint at optimum), the optimal strategy is to speed up local computing. However, with the increase of $\beta_1^E$, the WD1 prefers to save energy with a lower optimal $f^*_{i,1}$.
  \item For the $i$-th task of WD2, $i\in\{1,...,k-1\}$, a larger $\mu^*$ leads to a higher optimal $f_{i,2}^*$. On the other hand, the optimal $f_{i,2}^*$ is not related to $\mu^*$ for $i\in\{k,...,N\}$, as the corresponding executions are not constrained by the WDs' dependency.
\end{itemize}

\emph{\textbf{Proposition 3.2:}} $\forall i$ with $a_{i,1}=1$, the optimal transmit power of WD1 $p_{i,1}^*$ is expressed in \eqref{p1}, where $A_{1}=1+\frac{\beta_{1}^{T}+\lambda^{*}}{\beta_{1}^{E}P_{peak}}$, $B_{1}=\frac{h_{i,1}(\beta_{1}^{T}+\lambda^{*})}{\beta_{1}^{E}\sigma^{2}}-1$, $A_{2}=1+\frac{\lambda^{*}}{\beta_{1}^{E}P_{peak}}$ and $B_{2}=\frac{h_{i,1}\lambda^{*}}{\beta_{1}^{E}\sigma^{2}}-1$. Besides, $\forall i$ with $a_{i,2}=1$, the optimal transmit power of WD2 $p_{i,2}^*$ is expressed in \eqref{p2}, where $A_{3}=1+\frac{\beta_{2}^{T}}{\beta_{2}^{E}P_{peak}}$, $B_{3}=\frac{h_{i,2}\beta_{2}^{T}}{\beta_{2}^{E}\sigma^{2}}-1$, $A_{4}=1+\frac{\mu^{*}}{\beta_{2}^{E}P_{peak}}$ and $B_{4}=\frac{h_{i,2}\mu^{*}}{\beta_{2}^{E}\sigma^{2}}-1$.

Here, $\mathcal{W}(x)$ denotes the Lambert $\mathcal{W}$ function, which is the inverse function of $z\exp(z)=x$, i.e., $z=\mathcal{W}(x)$.

\begin{figure*}[htb]
\begin{small}
\begin{align}\label{p1}
\left\{
  \begin{array}{ll}
    \mbox{If} ~~~i\in\{1,...,M\}, & p_{i,1}^{*}=\left\{
                            \begin{array}{ll}
                              P_{peak}, & h_{i,1}<\frac{\sigma^{2}}{P_{peak}}\left[\frac{A_{1}}{-\mathcal{W}\left(-A_{1}e^{-A_{1}}\right)}-1\right], \\
                              \frac{\sigma^{2}}{h_{i,1}}\left[\frac{B_{1}}{\mathcal{W}\left(B_{1}e^{-1}\right)}-1\right], & otherwise.
                            \end{array}
                          \right.
\\
    \mbox{If} ~~~i=M+1, & p_{i,1}^{*}=\left\{
                            \begin{array}{ll}
                              P_{peak}, & h_{i,1}<\frac{\sigma^{2}}{P_{peak}}\left[\frac{A_{2}}{-\mathcal{W}\left(-A_{2}e^{-A_{2}}\right)}-1\right], \\
                              \frac{\sigma^{2}}{h_{i,1}}\left[\frac{B_{2}}{\mathcal{W}\left(B_{2}e^{-1}\right)}-1\right], & otherwise.
                            \end{array}
                          \right.
  \end{array}
\right.
\end{align}
\end{small}

\begin{small}
\begin{align}\label{p2}
\left\{
  \begin{array}{ll}
  \mbox{If} ~~~i\in\{1,...,k\}, & p_{i,2}^{*}=\left\{
                            \begin{array}{ll}
                              P_{peak}, & h_{i,2}<\frac{\sigma^{2}}{P_{peak}}\left[\frac{A_{4}}{-\mathcal{W}\left(-A_{4}e^{-A_{4}}\right)}-1\right], \\
                              \frac{\sigma^{2}}{h_{i,2}}\left[\frac{B_{4}}{\mathcal{W}\left(B_{4}e^{-1}\right)}-1\right], & otherwise.
                            \end{array}
                          \right.
\\
\mbox{If} ~~~i\in\{k+1,...,N\}, & p_{i,2}^{*}=\left\{
                            \begin{array}{ll}
                              P_{peak}, & h_{i,2}<\frac{\sigma^{2}}{P_{peak}}\left[\frac{A_{3}}{-\mathcal{W}\left(-A_{3}e^{-A_{3}}\right)}-1\right], \\
                              \frac{\sigma^{2}}{h_{i,2}}\left[\frac{B_{3}}{\mathcal{W}\left(B_{3}e^{-1}\right)}-1\right], & otherwise.
                            \end{array}
                          \right.
  \end{array}
\right.
\end{align}
\end{small}
\end{figure*}

\begin{proof}
Please refer to Appendix \ref{appendicesB}.
\end{proof}

From Proposition 3.2, we obtain the following observations:
\begin{itemize}
  \item The optimal transmit power is inversely proportional to the channel gain $h_{i,j}$ when $h_{i,j}$ is above a threshold, and equals the peak power $P_{peak}$ when the channel gain is below the threshold.
  \item With the increase of peak transmit power $P_{peak}$, the value of the threshold is decreasing. This means that for a larger $P_{peak}$, the WDs tend to transmit at the maximum power when meeting worse channel condition.
\end{itemize}

Based on Propositions 3.1 and 3.2, our precedent conference paper \cite{myglobecom} applies an ellipsoid method \cite{convex} to search for the optimal dual variables $(\lambda, \mu)$. The ellipsoid method guarantees to converge because (P2) is a convex problem given $\mathbf{a}$. In general, the ellipsoid method may take a long time to converge.

In this paper, we further study some interesting properties of an optimal solution in the following Lemma 3.1 and 3.2, based on which a reduced complexity one-dimensional bi-section search method is proposed in the following subsection.

\emph{\textbf{Lemma 3.1:}} $T^{wait}_1\leq T^{wait}_2$ and $\mu^*>0$ hold at the optimum of (P2).

\begin{proof}
We prove this lemma by contradiction. Suppose that there exists an optimal solution $\{\tau_{i,j}^{l},\tau_{i,j}^{u}\}$ with $T^{wait}_1> T^{wait}_2$. According to the KKT conditions $\lambda^{*}\left(T^{wait}_1-t\right)=0$ and $\mu^{*}\left(T^{wait}_2-t\right)=0$, we have $\lambda^*>0$ and $\mu^*=0$. As $\lambda^*>0$, according to \eqref{f1} and \eqref{p1}, the optimal $f_{i,1}^*$ and $p_{i,1}^*$ are finite, which means that $\{(\tau_{i,1}^{l})^*,(\tau_{i,1}^{u})^*\}$ are finite for all $i$. Hence, $T^{wait}_1$ is finite. However, when $\mu^*=0$, we have the optimal $(\tau_{i,2}^{l})^*\rightarrow\infty, i\in\{1,...,k-1\}$ from \eqref{f2} and $(\tau_{i,2}^{u})^*\rightarrow\infty, i\in\{1,...,k\}$ from \eqref{p2}. Thus, we have $T^{wait}_2\rightarrow\infty$. This contradicts with the assumption that $T^{wait}_1> T^{wait}_2$, and thus completes the proof.
\end{proof}

The above lemma indicates that the $k$-th task's waiting time for the input data stream $O_{M,1}$ from WD1 is not larger than that for the other input $O_{k-1,2}$ from WD2. In other words, WD2 always receives the task output from WD1 first and then waits until its local tasks finish before computing the $k$-th task.
In addition to the results in Lemma 3.1, the following lemma 3.2 shows two special cases, where $T^{wait}_1=T^{wait}_2$ is satisfied.

\emph{\textbf{Lemma 3.2:}} $T^{wait}_1=T^{wait}_2$ holds at the optimum of (P2) if one of the following two sufficient conditions is satisfied:
\begin{enumerate}
\item $\beta_1^T=0$;
\item $0<\beta_1^T<1$ and $a_{M,1}=0$.
\end{enumerate}

\begin{proof}
The proof is similar as that of Lemma 3.1 and is omitted here.
\end{proof}
Specifically, in the first case, the role of WD1 is solely to provide needed data to WD2 and minimizing its own execution time is not an objective. Nonetheless, the execution time of WD1 still affects that of WD2, which is to be minimized.  In the second case, the $M$-th task of WD1 chooses to perform local computing, i.e., $a_{M,1}=0$.





\subsection{A Low-complexity Bi-section Search Method}
According to Lemma 3.1, we have $t=\max\{T^{wait}_1,T^{wait}_2\}=T^{wait}_{2}$. Therefore, Problem (P2) is simplified as
\begin{eqnarray}
\mbox{(P3)}~~\min_{(\mathbf{a},\mathbf{p},\mathbf{f})}&&\eta_{1}+\eta_{2},\nonumber\\
{\rm s.t.}&&T^{wait}_1\leq T^{wait}_2,\nonumber\\
&&0\leq p_{i,j}\leq P_{peak},\nonumber\\
&&0\leq f_{i,j}\leq f_{peak},\nonumber\\
&&a_{i,j}\in\{0,1\},\forall i,j\nonumber.
\end{eqnarray}

Similarly, the Lagrangian of Problem (P3) is
\begin{align}
L'(\mathbf{p},\mathbf{f},\nu)&=\eta_{1}+\eta_{2}+\nu\left(T^{wait}_1-T^{wait}_2\right),
\end{align}
where $\nu\geq 0$ denotes the dual variable associated with the constraint $T^{wait}_1\leq T^{wait}_2$.

By applying the KKT conditions in (P3), we can obtain the optimal solutions of $\mathbf{f}$ and $\mathbf{p}$. The details are omitted here. By combining with the optimal solutions in Proposition 3.1 and Proposition 3.2, we have the following proposition.

\emph{\textbf{Proposition 3.3:}} The optimal dual variables $\{\lambda^*,\mu^*\}$ in (P2) and $\nu^*$ in (P3) are related by
\begin{align}\label{dual_relation}
\left\{
  \begin{array}{ll}
    \lambda^*=\nu^*,\\
    \mu^*=\beta_2^T-\nu^*,
  \end{array}
\right.
\end{align}
where $\nu^*\in[0,\beta_2^T)$. In other words, we have
\begin{align}
\lambda^*+\mu^*=\beta_2^T.
\end{align}

Note that (P3) is convex given the offloading decision $\mathbf{a}$. Thus, $\nu^*\left(T^{wait}_1-T^{wait}_2\right)=0$ is a sufficient condition for optimality. By defining $\psi(\nu)=T^{wait}_1-T^{wait}_2$, we can efficiently obtain the optimal $\nu^*$ based on the following proposition.

\emph{\textbf{Proposition 3.4:}} $\psi(\nu)$ is a monotonically decreasing function in $\nu\in[0,\beta_2^T)$. Besides, a unique $\nu^*\in(0,\beta_2^T)$ that satisfies $\psi(\nu^{*})=0$ exists when $\psi(\nu=0)>0$.
\begin{proof}
It can be proved that both $f_{i,1}(\nu)$ and $p_{i,1}(\nu)$ are monotonically increasing function in $\nu$, while $f_{i,2}(\nu)$ and $p_{i,2}(\nu)$, $1\leq i\leq k$, are monotonically decreasing function in $\nu$. Therefore, all terms in $\psi(\nu)$ decrease with $\nu$, thus $\psi(\nu)$ is a monotonically decreasing function in $\nu$. Meanwhile, when $\nu\rightarrow\beta_2^T$, it holds that $f_{i,2}(\nu)\rightarrow 0$ and $p_{i,2}(\nu)\rightarrow0$, $1\leq i\leq k$, which leads to $\psi(\nu)\rightarrow-\infty$ when $\nu\rightarrow\beta_2^T$. Together with the result that $\psi(\nu)$ is a monotonically decreasing function, there must exist a unique $\nu^*\in(0,\beta_2^T)$ that satisfies $\psi(\nu)=0$ when $\psi(\nu=0)>0$.
\end{proof}

\begin{algorithm}[tb]
\caption{Bi-section search method for Problem (P3) with given offloading decision $\mathbf{a}$}
\begin{algorithmic}[1]
\STATE \textbf{initialize} $\varepsilon=0.001$;
\STATE $\nu^{UB}=\beta_2^T$, $\nu^{LB}=0$;
\IF {$(T^{wait}_1-T^{wait}_2)|_{\nu=\nu^{LB}}<0$}
\STATE Set $\nu=\nu^{LB}$,$\lambda=\nu$, $\mu=\beta_2^T-\nu$;
\STATE Compute $\mathbf{f}$ according to \eqref{f1} and \eqref{f2};
\STATE Compute $\mathbf{p}$ according to \eqref{p1} and \eqref{p2};
\ELSE
\REPEAT
\STATE Set $\nu=\frac{\nu^{UB}+\nu^{LB}}{2}$, $\lambda=\nu$, $\mu=\beta_2^T-\nu$;
\STATE Compute $\mathbf{f}$ according to \eqref{f1} and \eqref{f2};
\STATE Compute $\mathbf{p}$ according to \eqref{p1} and \eqref{p2};
\IF{$T^{wait}_1-T^{wait}_2<0$}
\STATE $\nu^{UB}=\nu$
\ELSE
\STATE $\nu^{LB}=\nu$
\ENDIF
\UNTIL{$\left| T^{wait}_1-T^{wait}_2 \right|<\varepsilon$.}
\ENDIF
\end{algorithmic}
\end{algorithm}

With Proposition 3.4, when $\psi(\nu=0)>0$, the optimal $\nu^*$ can be efficiently obtained via a bi-section search over $\nu^*\in(0,\beta_2^T)$ that satisfies $\psi(\nu)=0$. If $\psi(\nu=0)<0$, we have $\nu^*=0$ according to the KKT condition $\nu^**\psi(\nu^*)=0$. Now that $\nu^*$ is obtained, the optimal $\{\mathbf{f}^*,\mathbf{p}^*\}$ can be directly calculated using \eqref{dual_relation}, \eqref{f1}, \eqref{f2}, \eqref{p1} and \eqref{p2}. Due to the convexity, the primal and dual optimal values are the same for (P3) given $\mathbf{a}$.

The pseudo-code of the bi-section search method is illustrated in Algorithm 1. Given a precision parameter $\varepsilon$, it takes $O\left(\log_{2}\left(\frac{\beta_2^T}{\varepsilon}\right)\right)$ number iterations for Algorithm 1 to converge. In each iteration, the computational complexity is proportional to the number of tasks in WDs, i.e., $O(M+N)$. Therefore, the overall complexity of Algorithm 1 is $O(M+N)$.

\section{Optimization of Offloading Decision}

In section III, we efficiently obtain the optimal $\{\mathbf{f},\mathbf{p}\}$ of (P1) once $\mathbf{a}$ is given. Intuitively, one can enumerate all $2^{M+N}$ feasible $\mathbf{a}$ and choose the optimal one that yields the minimum objective in (P2). However, this brute-force search quickly becomes computationally prohibitive as $(M+N)$ increases. In this section, we propose an efficient optimal Gibbs sampling algorithm to reduce the complexity.

\subsection{One-climb Policy}
Here, we first show in the following Theorem 1 that the optimal offloading decision $\mathbf{a}$ has an one-climb structure.

\begin{figure}
\begin{centering}
\includegraphics[scale=0.6]{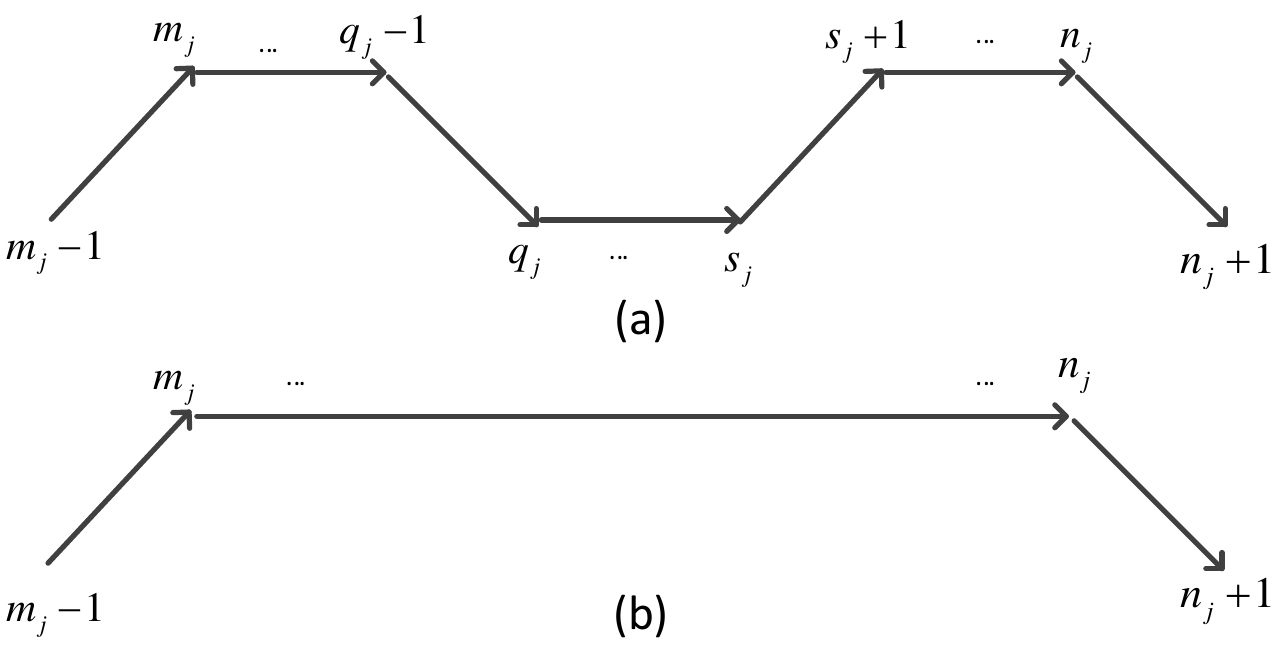}
\vspace{-0.1cm}
 \caption{ Illustration of a two-time offloading and an one-climb scenarios in WD $j$. }
\end{centering}
\vspace{-0.1cm}
\end{figure}

\emph{\textbf{Theorem 1 (one-climb policy):}} Assuming that $f_c>f_{peak}$, the execution for each WD migrates at most once from the WD to the edge server at the optimum.

 \begin{proof}
 In the following, we prove the one-climb policy by contradiction. Suppose that the optimal offloading decision allows a WD to offload its data more than one time, as shown in the Fig. 3(a). Under the two-time offloading scheme, tasks from $m_j$ to $q_j-1$ are migrated to the edge server for execution. Then, tasks from $q_j$ to $s_j$ execute at the WD $j$, followed by tasks from $s_j+1$ to $n_j$ migrated to the edge server, where $j$ is the index of WDs. As for the one-climb scheme in Fig. 3(b), tasks of WD $j$ from $q_j$ to $s_j$ are, however, executed on the edge server.

We denote the optimal offloading decision, local CPU frequencies and transmit power of WD $j$ in the two-time and one-climb offloading schemes as $\{\hat{\mathbf{a}}_j,\hat{\mathbf{f}}_j,\hat{\mathbf{p}}_j\}$ and $\{\tilde{\mathbf{a}}_j,\tilde{\mathbf{f}}_j,\tilde{\mathbf{p}}_j\}$, respectively.
%
By the optimality assumption, we have
\begin{align}
\eta_1(\hat{\mathbf{a}}_1,\hat{\mathbf{f}}_1,\hat{\mathbf{p}}_1)+\eta_2(\hat{\mathbf{a}}_2,\hat{\mathbf{f}}_2,\hat{\mathbf{p}}_2)<\eta_1(\tilde{\mathbf{a}}_1,\tilde{\mathbf{f}}_1,\tilde{\mathbf{p}}_1)+\eta_2(\tilde{\mathbf{a}}_2,\tilde{\mathbf{f}}_2,\tilde{\mathbf{p}}_2).
\end{align}

For the two-time offloading policy in WD1, the total execution time from the $m_1$-th task to the $n_1$-th task can be expressed as
\begin{align}
\hat{T}^{m_1\sim n_1}_{1}=&\sum_{i=m_1}^{q_1-1}(\tau_{i,1}^{c})+\tau_{q_1,1}^{d}+\sum_{i=q_1}^{s_1}(\tau_{i,1}^{l})\nonumber\\
&+\tau_{s_1+1,1}^{u}+\sum_{i=s_1+1}^{n_1}\tau_{i,1}^{c}.
\end{align}
As for the one-climb policy in WD1, we have
\begin{align}
\tilde{T}^{m_1\sim n_1}_{1}=\sum_{i=m_1}^{n_1}\tau_{i,1}^{c}.
\end{align}

Since the computing speed of the edge server is higher than that of the WDs, i.e., $f_c>f_{peak}$, the following inequalities hold for the $q_1$-th and $s_1$-th tasks:
\begin{align}
\tau_{q_1,1}^{c}<\tau_{q_1,1}^{l}<\tau_{q_1,1}^{l}+\tau_{q_1,1}^{d},
\end{align}
\begin{align}
\tau_{s_1,1}^{c}<\tau_{s_1,1}^{l}<\tau_{s_1,1}^{l}+\tau_{s_1+1,1}^{u}.
\end{align}
In addition, we have  $\tau_{i,1}^{c}<\tau_{i,1}^{l}, i=q_1,...,s_1$ for the tasks of WD1 between $q_1$ and $s_1$. Therefore, it can be shown that $\hat{T}^{m_1\sim n_1}_{1}>\tilde{T}^{m_1\sim n_1}_{1}$.

On the other hand, with respect to the energy consumption of WD1 from the $m_1$-th task to the $n_1$-th task, we observe that the two-time offloading scheme consumes more energy compared with the one-climb policy due to the local tasks computing $e_{i,1}^l$ from $q_1$ to $s_1$, the $(s_1+1)$-th task's offloading $e_{s_1+1,1}^u$ and the $M$-th task's offloading $e_{M+1,1}^u$ as illustrated in Fig. 2 (if $M\in\{q_1,...,s_1\}$). That is, $ \hat{E}^{m_1\sim n_1}_{1}>\tilde{E}^{m_1\sim n_1}_{1}$, where $\hat{E}^{m_1\sim n_1}_{1}$ and $\tilde{E}^{m_1\sim n_1}_{1}$ denote the energy consumption from the $m_1$-th task to the $n_1$-th task in the two-time and one-climb offloading schemes, respectively.

Similarly, as for the WD2, if $k\notin\{q_2,...,s_2\}$, $\hat{T}^{m_2\sim n_2}_{2}>\tilde{T}^{m_2\sim n_2}_{2}$ and $ \hat{E}^{m_2\sim n_2}_{2}>\tilde{E}^{m_2\sim n_2}_{2}$ hold according to the above discussion. Since extra time cost $\tau_{k',2}^d$ will be introduced if $a_{k,2}=0$ according to the task dependency model illustrated in Fig. 2, we still have $\hat{T}^{m_2\sim n_2}_{2}>\tilde{T}^{m_2\sim n_2}_{2}$ and $ \hat{E}^{m_2\sim n_2}_{2}>\tilde{E}^{m_2\sim n_2}_{2}$ when $k\in\{q_2,...,s_2\}$.

Therefore, for each WD $j$, we have
\begin{align}\label{OCPF4}
\hat{T}^{m_j\sim n_j}_{j}>\tilde{T}^{m_j\sim n_j}_{j},
\end{align}
and
\begin{align}\label{OCPF5}
\hat{E}^{m_j\sim n_j}_{j}>\tilde{E}^{m_j\sim n_j}_{j}.
\end{align}

We first consider the optimal solution $\{\hat{\mathbf{a}}_j,\hat{\mathbf{f}}_j,\hat{\mathbf{p}}_j\}$ in the two-time offloading scheme. According to Lemma 3.1, $T^{wait}_1\leq T^{wait}_2$ holds. Then, by switching the offloading decision $\hat{\mathbf{a}}_1$ to  $\tilde{\mathbf{a}}_1$ for WD1 and keeping the other variables unchanged, $E_1$, $T_1$ and $T_1^{wait}$ decrease according to \eqref{OCPF4} and \eqref{OCPF5}. Therefore, $T_1^{wait} \leq T_2^{wait}$ still holds at the solution $\{\tilde{\mathbf{a}}_1,\hat{\mathbf{a}}_2,\hat{\mathbf{f}}_j,\hat{\mathbf{p}}_j\}$, which leads to fixed $E_2$ and $T_2$.  Accordingly, we have
\begin{align}\label{OCPF1}
\eta_1(\hat{\mathbf{a}}_1,\hat{\mathbf{f}}_1,\hat{\mathbf{p}}_1)+\eta_2(\hat{\mathbf{a}}_2,\hat{\mathbf{f}}_2,\hat{\mathbf{p}}_2)>\eta_1(\tilde{\mathbf{a}}_1,\hat{\mathbf{f}}_1,\hat{\mathbf{p}}_1)+\eta_2(\hat{\mathbf{a}}_2,\hat{\mathbf{f}}_2,\hat{\mathbf{p}}_2).
\end{align}
Then, by further switching the offloading decision $\hat{\mathbf{a}}_2$ to  $\tilde{\mathbf{a}}_2$ for WD2 and keeping the other variables unchanged, $E_2$ and $T_2^{wait}$ do not increase according to \eqref{OCPF4} and \eqref{OCPF5}. Therefore, the term $T^{wait}=\max\{T_1^{wait},T_2^{wait}\}$ in \eqref{T2} is also not increasing, which leads to a non-increasing $T_2$. Thus, we have
\begin{align}\label{OCPF2}
\eta_1(\tilde{\mathbf{a}}_1,\hat{\mathbf{f}}_1,\hat{\mathbf{p}}_1)+\eta_2(\hat{\mathbf{a}}_2,\hat{\mathbf{f}}_2,\hat{\mathbf{p}}_2)>\eta_1(\tilde{\mathbf{a}}_1,\hat{\mathbf{f}}_1,\hat{\mathbf{p}}_1)+\eta_2(\tilde{\mathbf{a}}_2,\hat{\mathbf{f}}_2,\hat{\mathbf{p}}_2).
\end{align}
Furthermore, note that the optimal $\{\hat{\mathbf{f}}_j,\hat{\mathbf{p}}_j\}$ in a two-time offloading scheme is a feasible solution in the one-climb offloading scheme of (P1), which indicates that
\begin{align}\label{OCPF3}
\eta_1(\tilde{\mathbf{a}}_1,\hat{\mathbf{f}}_1,\hat{\mathbf{p}}_1)+\eta_2(\tilde{\mathbf{a}}_2,\hat{\mathbf{f}}_2,\hat{\mathbf{p}}_2)\geq\eta_1(\tilde{\mathbf{a}}_1,\tilde{\mathbf{f}}_1,\tilde{\mathbf{p}}_1)+\eta_2(\tilde{\mathbf{a}}_2,\tilde{\mathbf{f}}_2,\tilde{\mathbf{p}}_2).
\end{align}
Combining the above inequation \eqref{OCPF1}, \eqref{OCPF2}, \eqref{OCPF3}, we have
\begin{align}
\eta_1(\hat{\mathbf{a}}_1,\hat{\mathbf{f}}_1,\hat{\mathbf{p}}_1)+\eta_2(\hat{\mathbf{a}}_2,\hat{\mathbf{f}}_2,\hat{\mathbf{p}}_2)>\eta_1(\tilde{\mathbf{a}}_1,\tilde{\mathbf{f}}_1,\tilde{\mathbf{p}}_1)+\eta_2(\tilde{\mathbf{a}}_2,\tilde{\mathbf{f}}_2,\tilde{\mathbf{p}}_2).
\end{align}

Therefore, it contradicts the assumption. Thus, for each WD, the one-climb policy is better than two-time offloading scheme. Similarly, the same conclusion can be drawn by comparing the one-climb policy with a $\phi$-time offloading scheme, where $\phi\geq 2$. It completes the proof.
%
 \end{proof}

The one-climb policy indicates that each WD either offloads its data only once to the edge server or does not offload at all at the optimum. Therefore, we only need to enumerate the offloading decisions that satisfy the one-climb policy, instead of all the $2^{M+N}$ feasible offloading decisions (as in the precedent conference paper \cite{myglobecom}). Specifically, under the one-climb policy, if task offloading is necessary, we only need to search for the two tasks of each WD, i.e., the tasks that data is offloaded to and downloaded from the AP, respectively. For WD1, we need to search $(((M+1)M)/2)+1$ such combinations of tasks, including the special case that the WD does not offload throughout the execution time. Similarly, WD2 has $(((N+1)N)/2)+1$ such combinations to search. Therefore, the total search space  is $[(((M+1)M)/2)+1][(((N+1)N)/2)+1]$, i.e., $\mathcal{O}(M^2\cdot N^2)$, which is significantly lower than the brute-force based method when $M$ or $N$ is larger. Table I illustrates the number of searches performed by the one-climb based scheme and the brute-force method under different $M$ and $N$.

Nonetheless, the proposed searching method may still induce high computational complexity when $M$ or $N$ is large. In the following, we further propose a reduced-complexity Gibbs sampling algorithm to optimize the offloading decisions.

\begin{table}
\centering
\begin{tabular}{cccc}
\toprule
 & One-climb based & Brute-force & $\Psi_1/\Psi_2$\\
  &scheme $\Psi_1$& method $\Psi_2$&  \\
\midrule
$M=5,N=10$& 896& $2^{15}$&$2.734\%$  \\
$M=10,N=10$& 3136& $2^{20}$& $0.299\%$  \\
$M=10,N=20$& 11816& $2^{30}$& $0.001\%$ \\
\bottomrule
\end{tabular}
\caption{The number of searches performed by the one-climb based scheme and the brute-force method under different $M$ and $N$.}
\end{table}

\subsection{One-climb Policy based Gibbs Sampling}

Gibbs Sampling was originally introduced to model the physical interactions between molecules and particles. There are many modern engineering applications of Gibbs sampling, e.g., on image processing in \cite{GS1} and nonconvex power control in \cite{GS2}. Specifically, Gibbs Sampling solves an optimization problem with the following form:
\begin{align}
\min_{\mathbf{x}\in \mathcal{X}}J(\mathbf{x}),
\end{align}
where the variable $\mathbf{x}$ is a $D$-dim row vector with element $x_d$, $d=1,...,D$, and the objective function $J(\mathbf{x})$ can be of any form.

In Gibbs Sampling, the value of each $x_d$ is updated iteratively and asynchronously according to the probability distribution $\mathbf{\Lambda}_d(\mathbf{x}_{-d})=(\Lambda_d(x_d|\mathbf{x}_{-d}), \forall x_d\in\mathcal{X}_d)$ with
\begin{align}\label{gibbs prob}
\Lambda_d(x_d|\mathbf{x}_{-d})=\frac{\exp\left(-J\left(x_d,\mathbf{x}_{-d}\right)/T\right)}{\sum_{x'_d\in\mathcal{X}_d}\exp\left(-J\left(x'_d,\mathbf{x}_{-d}\right)/T\right)},
\end{align}
where $\mathbf{x}_{-d}=(x_1,...,x_{d-1},x_{d+1},...,x_D)$ and $T>0$ denotes the temperature parameter. According to \eqref{gibbs prob}, a $x_d$ that yields a better objective function value (i.e., a smaller $J(\cdot)$ here) will be picked with a higher probability. This is especially true when $T$ is small. According to the proof in [Section IV, 19], a Gibbs sampling algorithm obtains the optimal solution when it converges.

\begin{figure}
\begin{centering}
\includegraphics[scale=0.52]{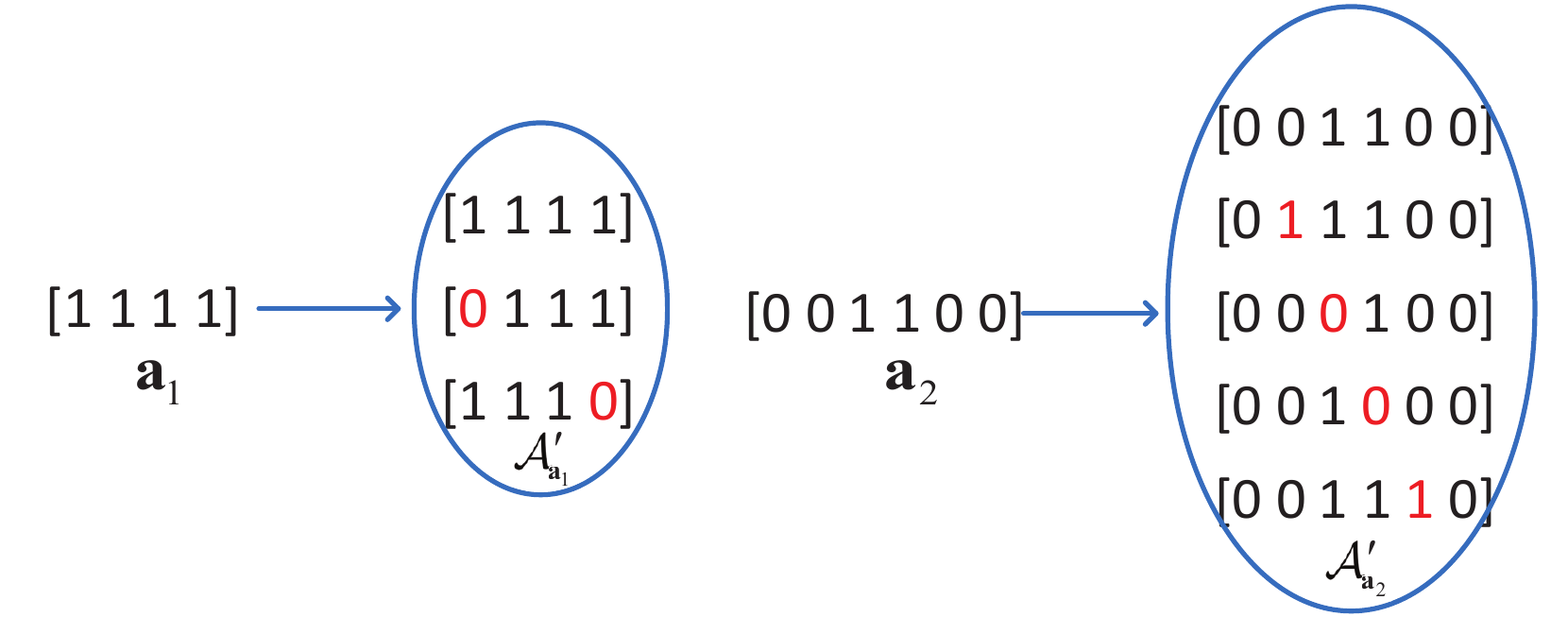}
\vspace{-0.1cm}
 \caption{ An example of the generating rule for the sampling sets $\mathcal{A}'_{\mathbf{a}_1}$ and $\mathcal{A}'_{\mathbf{a}_2}$ with given $\mathbf{a}_1=[1 1 1 1]$ and $\mathbf{a}_2=[0 0 1 1 0 0]$. }
\end{centering}
\vspace{-0.1cm}
\end{figure}

\begin{algorithm}[htb]
\caption{The proposed Gibbs sampling algorithm }
\begin{algorithmic}[1]
\STATE \textbf{initialize} $\mathbf{a}_1^{(0)}$, $\mathbf{a}_2^{(0)}$, $T(1)$ and $\theta=1$.
\REPEAT
\STATE Generate $\mathcal{A}'_{\mathbf{a}_1^{(\theta-1)}}$ and $\mathcal{A}'_{\mathbf{a}_2^{(\theta-1)}}$.
\STATE Sample $\mathbf{a}_1^{(\theta)}$ according to \eqref{GS1}.
\STATE Sample $\mathbf{a}_2^{(\theta)}$ according to \eqref{GS2}.
\STATE Set $\theta=\theta+1$ and $T(\theta)=\alpha T(\theta-1)$.
 \UNTIL{the optimal objective value of (P1) $\varphi$ converges.}
\end{algorithmic}
\end{algorithm}

\begin{figure*}
\begin{centering}
\includegraphics[scale=0.6]{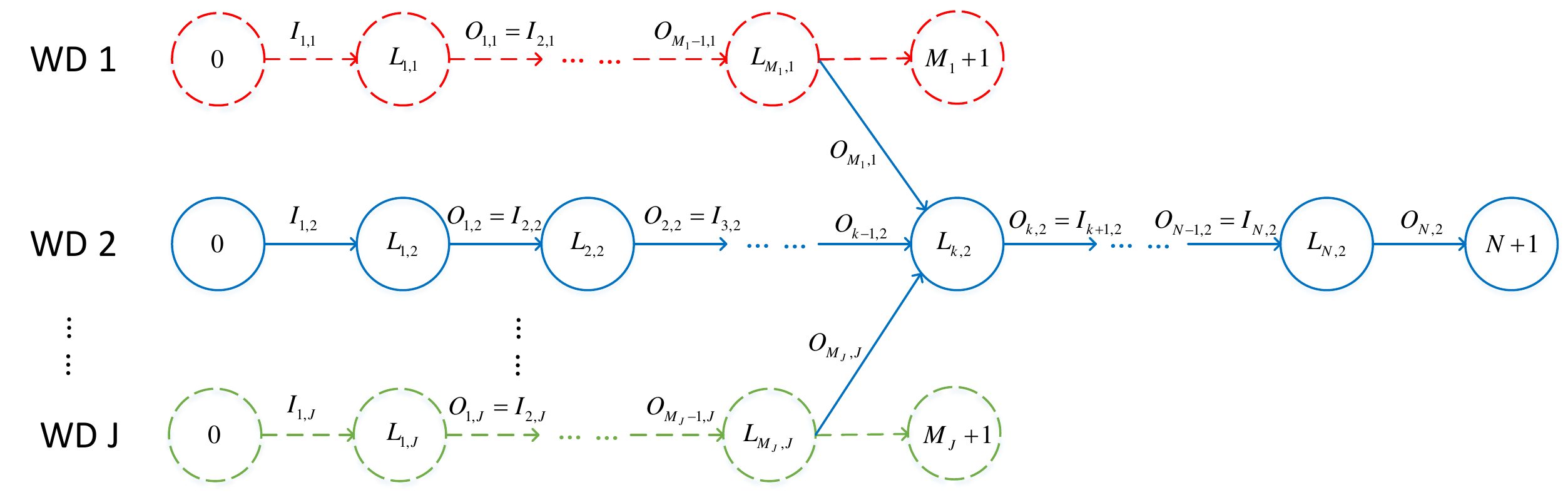}
\vspace{-0.1cm}
 \caption{ Extended inter-user task dependency model where the input of the task $k$ at WD2 requires the final task outputs from the other WDs. }
\end{centering}
\vspace{-0.1cm}
\end{figure*}

In our problem, we denote the offloading decisions corresponding to the WD1 and WD2 by two vectors, $\mathbf{a}_1$ and $\mathbf{a}_2$, respectively that satisfy the one-climb policy. $\varphi(\mathbf{a}_1,\mathbf{a}_2)$ denotes the optimal objective value of (P1) given $\mathbf{a}_1$ and $\mathbf{a}_2$. In addition, $\mathcal{A}'_{\mathbf{a}_1}$ and $\mathcal{A}'_{\mathbf{a}_2}$ denote the sampling sets generated from $\mathbf{a}_1$ and $\mathbf{a}_2$, respectively. The generating rule is that with given $\mathbf{a}_1$ and $\mathbf{a}_2$, at most one entry of $\mathbf{a}_1$ and $\mathbf{a}_2$ swaps from 1 to 0 (or 0 to 1) while the newly generated vector still satisfies the one-climb policy for each WD. Thus, there only exist a small number of feasible elements in $\mathcal{A}'_{\mathbf{a}_1}$ and $\mathcal{A}'_{\mathbf{a}_2}$. In Fig. 4, we show an example to illustrate the generating rule for the sampling sets $\mathcal{A}'_{\mathbf{a}_1}$ and $\mathcal{A}'_{\mathbf{a}_2}$ with given $\mathbf{a}_1$ and $\mathbf{a}_2$. Compared to Gibbs sampling without the one-climb policy, $40\%$ and $28.57\%$ lower search spaces corresponding to $\mathbf{a}_1$ and $\mathbf{a}_2$ can be achieved, respectively in the example of Fig. 4.

In one-climb policy based Gibbs sampling, the offloading decisions of WD1 and WD2 are updated to $\mathbf{a}_1^{(\theta)}$ and $\mathbf{a}_2^{(\theta)}$ in the $\theta$-th sampling according to the probability distributions $\mathbf{\Lambda}_1(\mathbf{a}_1|\mathbf{a}_2^{(\theta-1)})=\{\Lambda_1(\mathbf{a}_1|\mathbf{a}_2^{(\theta-1)}),\forall\mathbf{a}_1\in\mathcal{A}'_{\mathbf{a}_1^{(\theta-1)}}\}$ and $\mathbf{\Lambda}_2(\mathbf{a}_2|\mathbf{a}_1^{(\theta)})=\{\Lambda_2(\mathbf{a}_2|\mathbf{a}_1^{(\theta)}),\forall\mathbf{a}_2\in\mathcal{A}'_{\mathbf{a}_2^{(\theta-1)}}\}$ with
\begin{align}\label{GS1}
\Lambda_1(\mathbf{a}_1|\mathbf{a}_2^{(\theta-1)})=\frac{\exp\left(-\varphi(\mathbf{a}_1,\mathbf{a}_2^{(\theta-1)})/T\right)}{\sum_{\mathbf{a}'_1\in\mathcal{A}'_{\mathbf{a}_1^{(\theta-1)}}}\exp\left(-\varphi(\mathbf{a}'_1,\mathbf{a}_2^{(\theta-1)})/T\right)}
\end{align}
and
\begin{align}\label{GS2}
\Lambda_2(\mathbf{a}_2|\mathbf{a}_1^{(\theta)})=\frac{\exp\left(-\varphi(\mathbf{a}_2,\mathbf{a}_1^{(\theta)})/T\right)}{\sum_{\mathbf{a}'_2\in\mathcal{A}'_{\mathbf{a}_2^{(\theta-1)}}}\exp\left(-\varphi(\mathbf{a}'_2,\mathbf{a}_1^{(\theta)})/T\right)},
\end{align}
respectively. According to \eqref{GS1} and \eqref{GS2}, $\mathbf{a}_1$ or $\mathbf{a}_2$ that yields a smaller objective function value will be picked with a higher probability. However, one difficulty is that when $T$ is very small, the time it takes to reach equilibrium can be excessive \cite{GS3}. This drawback can be overcome by using a slowly decreasing ``cooling schedule" $T(\theta)=\alpha T(\theta-1)$, where $\alpha<1$ is the cooling rate. The pseudo-code of one-climb policy based Gibbs sampling algorithm is shown in Algorithm 2.

\section{The Multiuser Scenario}

In this section, we extend the proposed inter-user task dependency model consisting of only two users to a general multi-user case, where the input of a task at one WD requires the final task outputs from multiple other WDs. We assume that  there are $J$ WDs. As shown in Fig. 5, the calculation of the intermediate task $k$ of WD2 requires the final task outputs from the other $J-1$ WDs. Specifically, for WD2, we have $I_{k,2}=O_{k-1,2}+\sum_{j\neq 2}O_{M_j,j}$, where $M_j$ is the number of sequential tasks to execute at WD $j$, for $j\neq 2$.

In this case, the waiting time until the output of the $M_j$-th task of WD $j$ $(j\neq 2)$ reaches WD2, denoted by $T_{j}^{wait}$, can be expressed as
\begin{align}
T^{wait}_{j}=&\sum_{i=1}^{M_j}\bigg[(1-a_{i,j})\tau_{i,j}^{l}+a_{i,j}(\tau_{i,j}^{c}+\tau_{i,j}^{u})\nonumber\\
&+a_{i-1,j}\tau_{i,j}^{d}-a_{i-1,j}a_{i,j}(\tau_{i,j}^{u}+\tau_{i,j}^{d})\bigg]
\nonumber\\&+(1-a_{M_j,j})\tau_{M_j+1,j}^{u}+(1-a_{k,2})\frac{O_{M_j,j}}{R_{k,2}^{d}}.
\end{align}
Therefore, the total waiting time before the joint task is ready for execution in (19) becomes $T^{wait}=\max\{T^{wait}_{1},T^{wait}_{2},...,T^{wait}_{j},...,T^{wait}_{J}\}$.
We omit some details on formulation due to the page limit and rewrite the optimization problem (P2) in such multi-user task dependency model as
\begin{eqnarray}
\mbox{(P4)}~~\min_{(\mathbf{a},\{\tau_{i,j}^{u}\},\{\tau_{i,j}^{l}\},t)}&&\sum_{j=1}^{J}\eta_{j},\nonumber\\
{\rm s.t.}&&t\geq T^{wait}_{1},t\geq T^{wait}_2,...,\nonumber\\
&&t\geq T^{wait}_j,...,t\geq T^{wait}_J,\nonumber\\
&&\tau_{i,j}^{u}\geq\frac{O_{i-1,j}}{W\log_{2}\left(1+\frac{P_{peak}h_{i,j}}{\sigma^{2}}\right)},\nonumber\\
&&\tau_{i,j}^l\geq\frac{L_{i,j}}{f_{peak}},\nonumber\\
&&a_{i,j}\in\{0,1\},\forall i,j.\nonumber
\end{eqnarray}

\emph{\textbf{Lemma 5.1:}} $\forall j (j\neq 2)$, $T^{wait}_j\leq T^{wait}_2$ holds at the optimum.
\begin{proof}
The proof follows a similar technique in Lemma 3.1 by analyzing the KKT conditions of (P4), which is omitted due to the page limit.
\end{proof}

\emph{\textbf{Lemma 5.2:}} The optimal offloading decisions in the extended inter-user task dependency model follow the one-climb policy.
\begin{proof}
We prove the one-climb policy by contradiction based on the proof of Theorem 1. Suppose that the optimal offloading strategy is to offload each WD's data more than one time, then we have
\begin{align}
\sum_{j=1}^{J}\eta_j(\hat{\mathbf{a}}_j,\hat{\mathbf{f}}_j,\hat{\mathbf{p}}_j)<\sum_{j=1}^{J}\eta_j(\tilde{\mathbf{a}}_j,\tilde{\mathbf{f}}_j,\tilde{\mathbf{p}}_j).
\end{align}
According to the proof of Theorem 1, \eqref{OCPF4} and \eqref{OCPF5} also hold in the multi-user task dependency model. We first consider the optimal solution $\{\hat{\mathbf{a}}_j,\hat{\mathbf{f}}_j,\hat{\mathbf{p}}_j\}$ in the two-time offloading scheme. According to Lemma 5.1, $\forall j (j\neq 2)$, $T^{wait}_j\leq T^{wait}_2$ holds at the optimum. Then, we switch the offloading decision $\hat{\mathbf{a}}_j$ to  $\tilde{\mathbf{a}}_j$ for WD $j$ $(j\neq 2)$ successively and keep the other variables unchanged. Based on the analysis of \eqref{OCPF1} in the proof of Theorem 1, we have
\begin{align}
\sum_{j=1}^{J}\eta_j(\hat{\mathbf{a}}_j,\hat{\mathbf{f}}_j,\hat{\mathbf{p}}_j)>\sum_{j\neq 2}\eta_j(\tilde{\mathbf{a}}_j,\hat{\mathbf{f}}_j,\hat{\mathbf{p}}_j)+\eta_2(\hat{\mathbf{a}}_2,\hat{\mathbf{f}}_2,\hat{\mathbf{p}}_2).
\end{align}
Then, we further switch the offloading decision $\hat{\mathbf{a}}_2$ to  $\tilde{\mathbf{a}}_2$ for WD2 and keep the other variables unchanged. Based on the analysis of \eqref{OCPF2} in the proof of Theorem 1, we have
\begin{align}
\sum_{j\neq 2}\eta_j(\tilde{\mathbf{a}}_j,\hat{\mathbf{f}}_j,\hat{\mathbf{p}}_j)+\eta_2(\hat{\mathbf{a}}_2,\hat{\mathbf{f}}_2,\hat{\mathbf{p}}_2)>\sum_{j=1}^{J}\eta_j(\tilde{\mathbf{a}}_j,\hat{\mathbf{f}}_j,\hat{\mathbf{p}}_j).
\end{align}
Since the optimal $\{\hat{\mathbf{f}}_j,\hat{\mathbf{p}}_j\}$ in a two-time offloading scheme is a feasible solution in the one-climb offloading scheme of (P4), we have
\begin{align}
\sum_{j=1}^{J}\eta_j(\tilde{\mathbf{a}}_j,\hat{\mathbf{f}}_j,\hat{\mathbf{p}}_j)\geq\sum_{j=1}^{J}\eta_j(\tilde{\mathbf{a}}_j,\tilde{\mathbf{f}}_j,\tilde{\mathbf{p}}_j).
\end{align}
Therefore, by combining the above three inequalities, we have
\begin{align}
\sum_{j=1}^{J}\eta_j(\hat{\mathbf{a}}_j,\hat{\mathbf{f}}_j,\hat{\mathbf{p}}_j)>\sum_{j=1}^{J}\eta_j(\tilde{\mathbf{a}}_j,\tilde{\mathbf{f}}_j,\tilde{\mathbf{p}}_j),
\end{align}
which contradicts the assumption. It completes the proof.
\end{proof}

According to Lemma 5.2, the proposed reduced-complexity Gibbs sampling algorithm can
be adapted to solve the problem. Specifically, for the $\theta$-th iteration, we first generate the sampling set $\mathcal{A}'_{\mathbf{a}_j^{(\theta-1)}}$ for each WD $j$ based on the one-climb policy. Then, the offloading decision $\mathbf{a}_j^{(\theta)}$ of each WD $j$ is sampled sequentially according to the probability distribution $\mathbf{\Lambda}_j(\mathbf{a}_j|\mathbf{a}_{-j})$, which is similar as \eqref{GS1} and \eqref{GS2}. We omit the details here due to the page limit.

In this paper, we assume that the edge server is equipped with $\rho$ cores and each core is assigned to compute one task with a fixed service rate $f_c$.  Since each WD has a sequence of tasks to execute, there are at most $J$ tasks executed at the edge simultaneously. Therefore, the maximum acceptable number of users is $\rho$ and $J\leq\rho$ must hold.

\section{Numerical Results}

\begin{figure}
\begin{centering}
\includegraphics[scale=0.45]{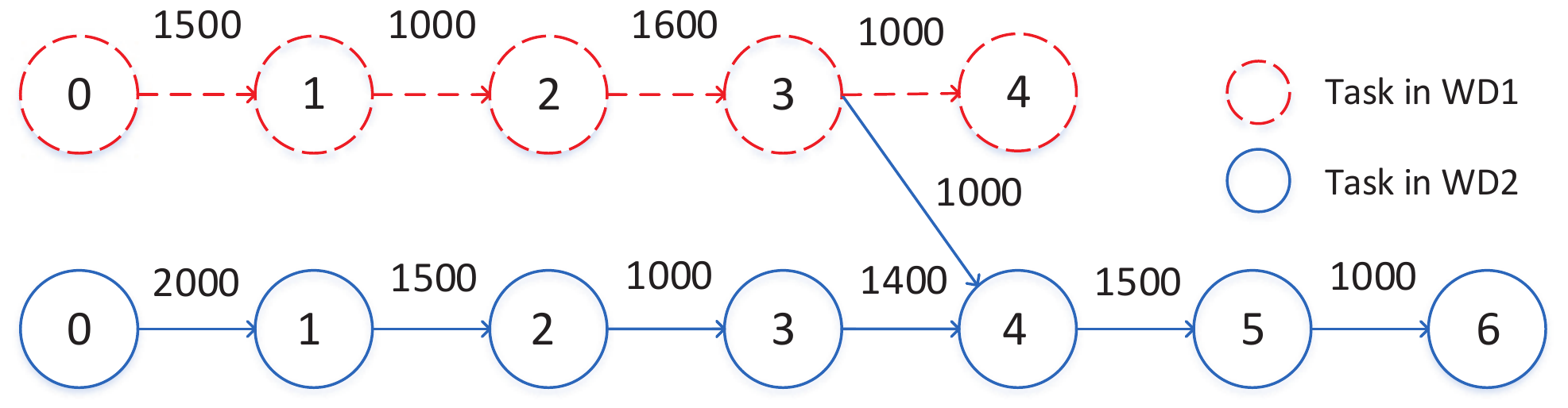}
\vspace{-0.1cm}
 \caption{ The considered topological call graph in simulation. }\label{example}
\end{centering}
\vspace{-0.1cm}
\end{figure}


In this section, we conduct numerical simulations to evaluate the performance of our optimal strategies. Consider an example call graph in Fig. \ref{example}. The input and output data size (KByte) of each task are shown in Fig. \ref{example}. As for the computing workload, we assume that $\{L_{i,1}\}=[65.5 ~40.3~96.6]$ (Mcycles) and $\{L_{i,2}\}=[70.8 ~95.3 ~86.4 ~18.6 ~158.6]$ (Mcycles).  We assume that the transmit power at the AP is fixed as 1 W and the peak transmit power of each WD is 100 mW. Besides, the edge server speed $f_c$ and the peak computational frequency of each WD $f_{peak}$ are equal to $10^{10}$ and $10^8$ cycles/s, respectively. We consider a commercial mobile device in practice with the computing efficiency parameter $\kappa=10^{-26}$, which is consistent with the measurements in \cite{parameter1}.

For simplicity of illustration, we assume that the wireless channel gains $h_{i,j}, g_{i,j}$ follow the free-space path loss model
\begin{align}
h_{i,j}=g_{i,j}=G\left(\frac{3\cdot 10^8}{4\pi F_cd_j}\right)^{PL},
\end{align}
where $G=4.11$ denotes the antenna gain, $F_c=915$ MHz denotes the carrier frequency, $d_j$ in meters denotes the distance between the WD $j$ and the AP, and $PL=3$ denotes the path loss exponent. In this case, the wireless channel gains are equal for all the tasks at a WD. However, our proposed algorithms are applicable to general cases with different $h_{i,j}, g_{i,j}$ for different task $i$ of WD $j$. The noise power $\sigma^2=10^{-10}$ W. We set the bandwidth $W=2$ MHz. Recall that the weights in WD $j$ are related by $\beta_j^E=1-\beta_j^T$.  In general, the parameters chosen in the simulation are based on practical computing models \cite{parameter1} and typical wireless networks \cite{parameter2}.


\begin{figure}
\begin{centering}
\includegraphics[scale=0.6]{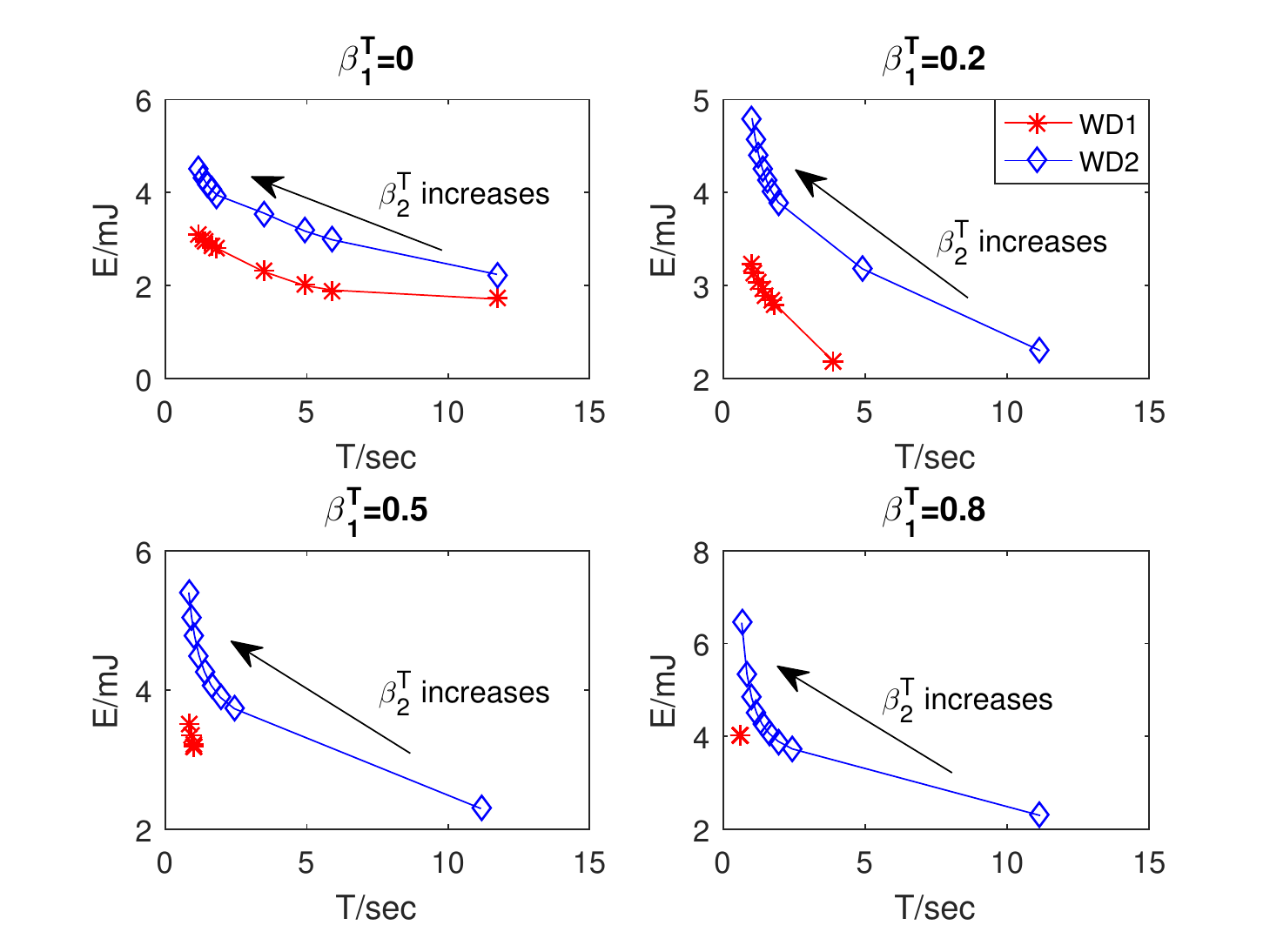}
\vspace{-0.1cm}
 \caption{ The tradeoff between the total execution time and energy consumption of each WD when $\beta_2^T$ varies. }
\end{centering}
\vspace{-0.1cm}
\end{figure}

\subsection{Energy Efficiency and Delay Performance Evaluation}

In Fig. 7, we study the performance tradeoff between energy consumption and delay for the two WDs under different $\beta_1^T$ and $\beta_2^T$. Here, we consider $d_1=d_2=15$ m. Under each particular $\beta_1^T$, it can be seen that with the increase of $\beta_2^T$, WD2 achieves lower execution delay but higher total energy consumption. Similar performance tradeoff is also observed for WD1. Moreover, we observe that the tradeoff curve of WD1 converges to a point as $\beta_1^T$ increases, which means that for a large $\beta_1^T$, the optimal execution time and energy consumption of WD1 remains constant with the increase of $\beta_2^T$. It is due to the fact that with the increase of $\beta_1^T$, the WD1 not only acts as a helper, but also focuses on minimizing its own execution time.

Then, we show the ETC objective value achieved by different methods when $d_1$ and $d_2$ varies, where we set $\beta_1^T=0.05$ and $\beta_2^T=0.5$. For performance comparison, we also consider three suboptimal schemes as benchmarks. The first scheme is referred to as all task offloading, where all the tasks in the two WDs are offloaded to the edge. For the second scheme, all the tasks of the two WDs are executed locally. Besides, we denote independent optimization as the third scheme, where each WD minimizes its own ETC and neglects the task dependency between them.



\begin{figure}
\begin{centering}
\includegraphics[scale=0.6]{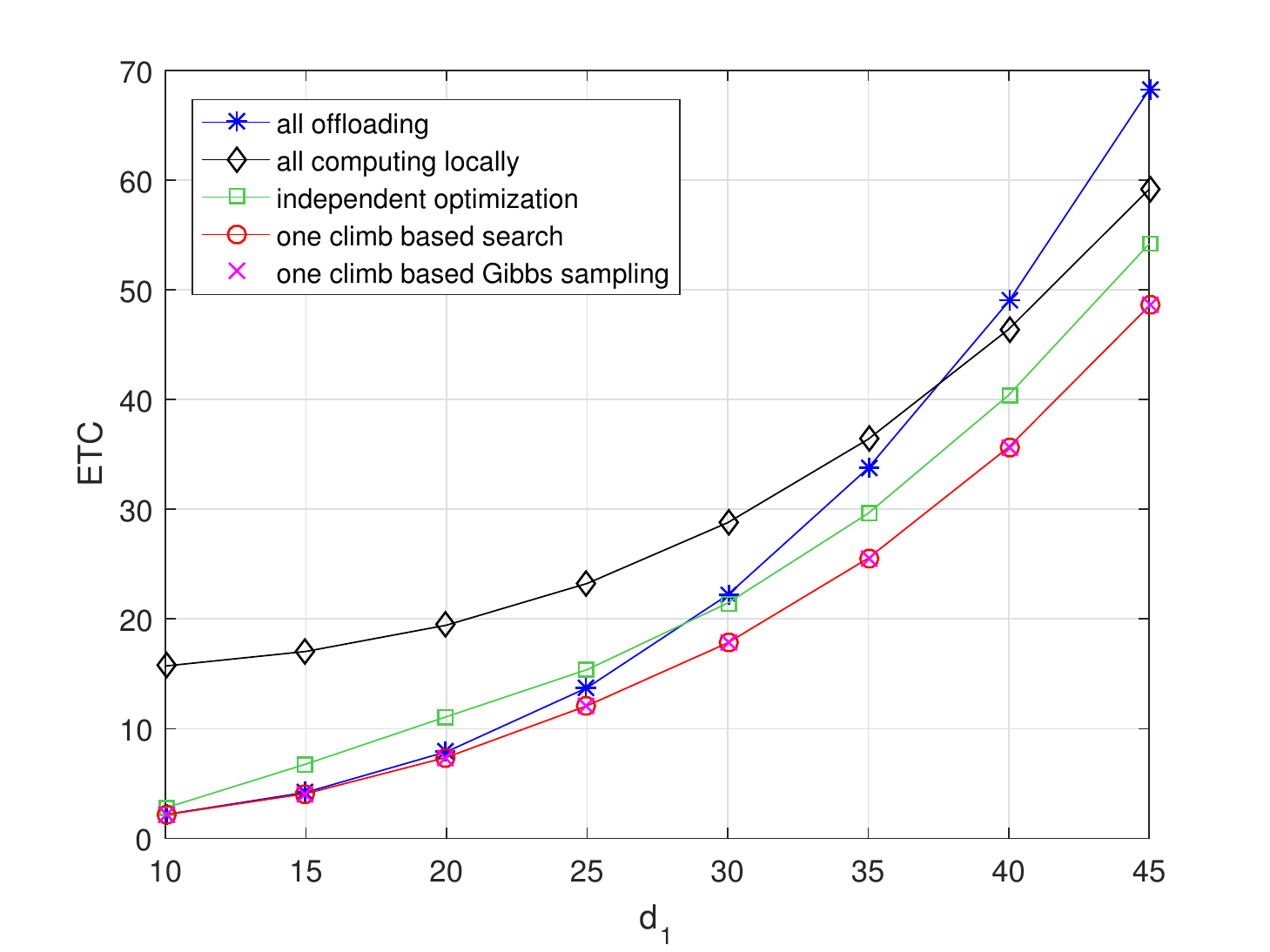}
\vspace{-0.1cm}
 \caption{ Total ETC versus $d_1$ when $d_2=10$ m. }
\end{centering}
\vspace{-0.1cm}
\end{figure}

\begin{figure}
\begin{centering}
\includegraphics[scale=0.6]{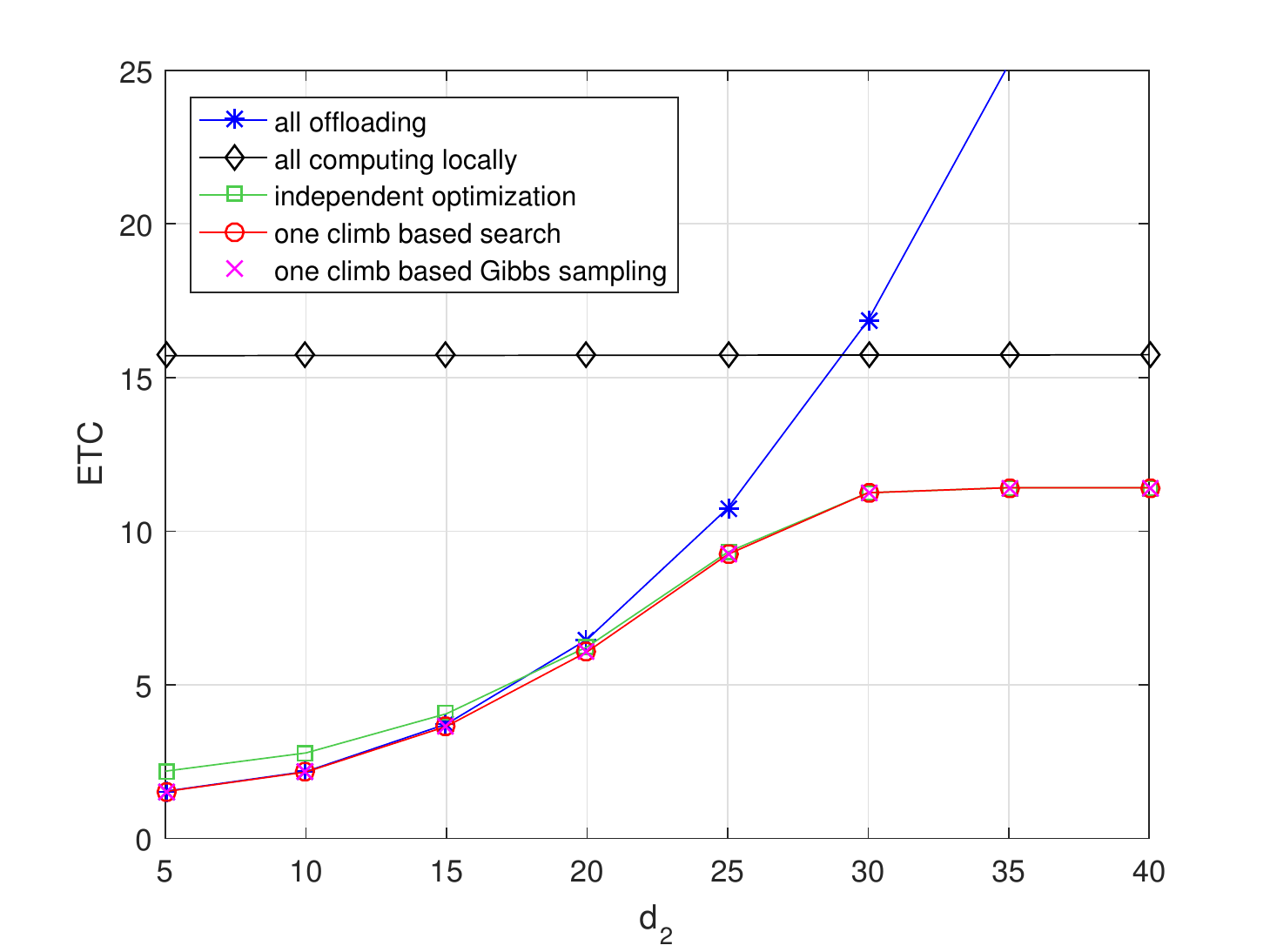}
\vspace{-0.1cm}
 \caption{ Total ETC versus $d_2$ when $d_1=10$ m. }
\end{centering}
\vspace{-0.1cm}
\end{figure}

In Fig. 8, we illustrate the impact of $d_1$ on the total ETC, where $d_2$ is fixed as 10 m. Besides, Fig. 9 demonstrates the total ETC when $d_2$ varies with $d_1=10$ m. The $\{L_{i,1}\}$ and $\{L_{i, 2}\}$ are uniformly generated from the range $[10,200]$ (Mcycles). Each point in the figures is the average performance of 20 independent simulations. From both figures, it can be seen that the optimal ETC obtained by the proposed Gibbs sampling algorithm is on top of each other with the optimal one-climb policy based enumeration method. In addition, it is observed from both Fig. 8 and Fig. 9 that the total ETC is increasing as $d_1$ or $d_2$ increases for the proposed algorithm, all-offloading scheme and independent optimization scheme. As for the all-computing-locally scheme, higher total ETC is achieved with the increase of $d_1$, while the total ETC is more stable when $d_2$ increases. It is because in the all-computing-locally scheme, the WD1 needs to upload its final result to the AP and then, the AP forwards this information to the WD2, as illustrated in Fig. 2 Case1. In this process, increasing $d_1$ leads to a higher total ETC.
Besides, it is observed that lower ETC is achieved by the proposed algorithm compared to the three benchmarks, i.e., around $15.24\%$, $47.64\%$ and $21.2\%$ lower average ETC than the all-offloading, all-computing-locally and independent optimization schemes in Fig. 8, respectively. This suggests the benefits by adapting joint optimization of the resource allocation and the offloading decisions for both WDs. An interesting observation is that the independent optimization scheme performs
equally well as the proposed optimal algorithm with larger $d_2$ in Fig. 9. It is due to the fact that $T_{1}^{wait}<T_{2}^{wait}$ and $\lambda^*=0$ when $d_2$ is large in the proposed scheme. This implies that the optimizations of the two WDs are practically decoupled.


\begin{figure}
\begin{centering}
\includegraphics[scale=0.6]{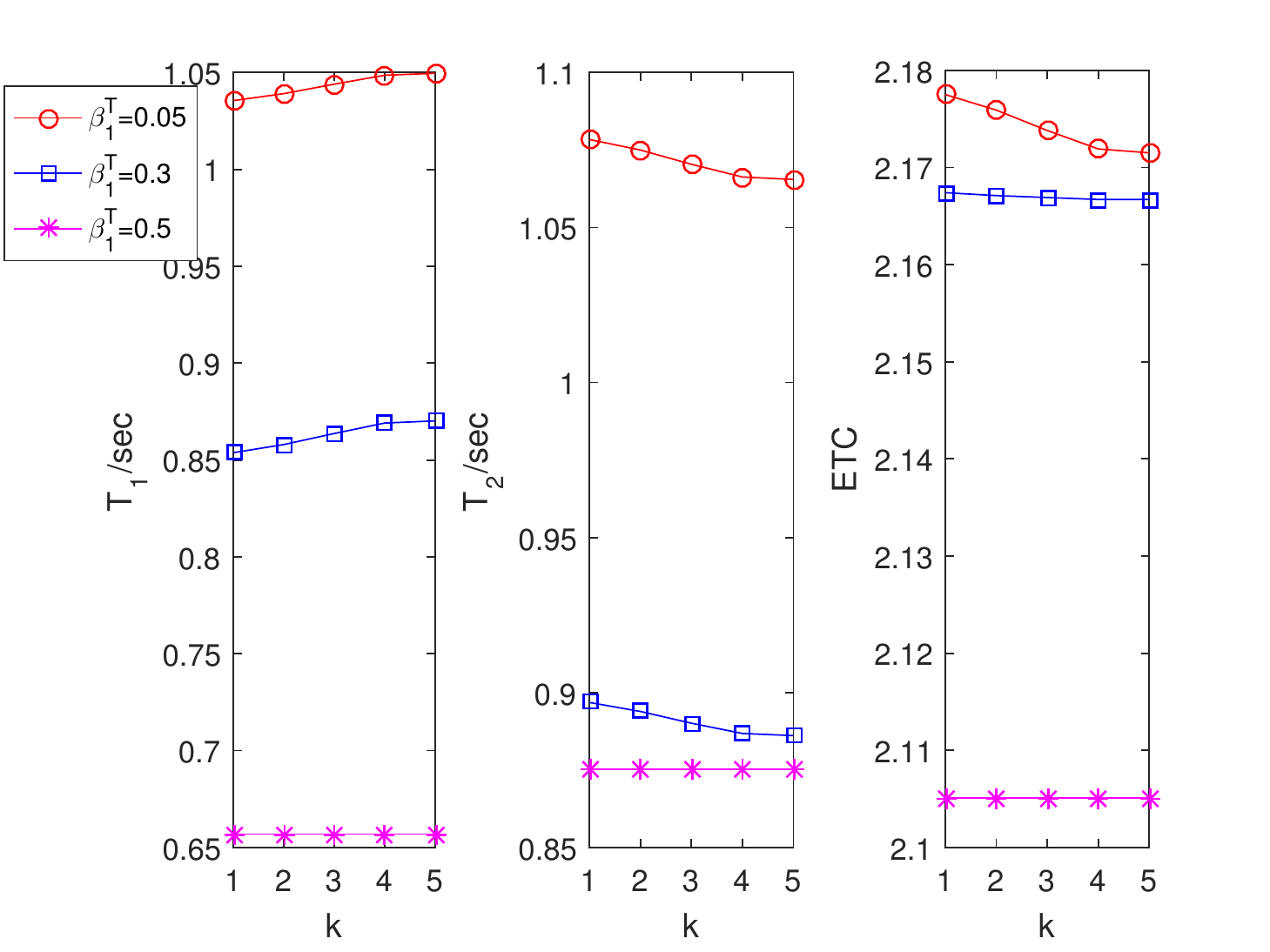}
\vspace{-0.1cm}
 \caption{ Impact of $k$ in topological call graphs $(3,5,k)$. }
\end{centering}
\vspace{-0.1cm}
\end{figure}

\begin{figure}
\begin{centering}
\includegraphics[scale=0.6]{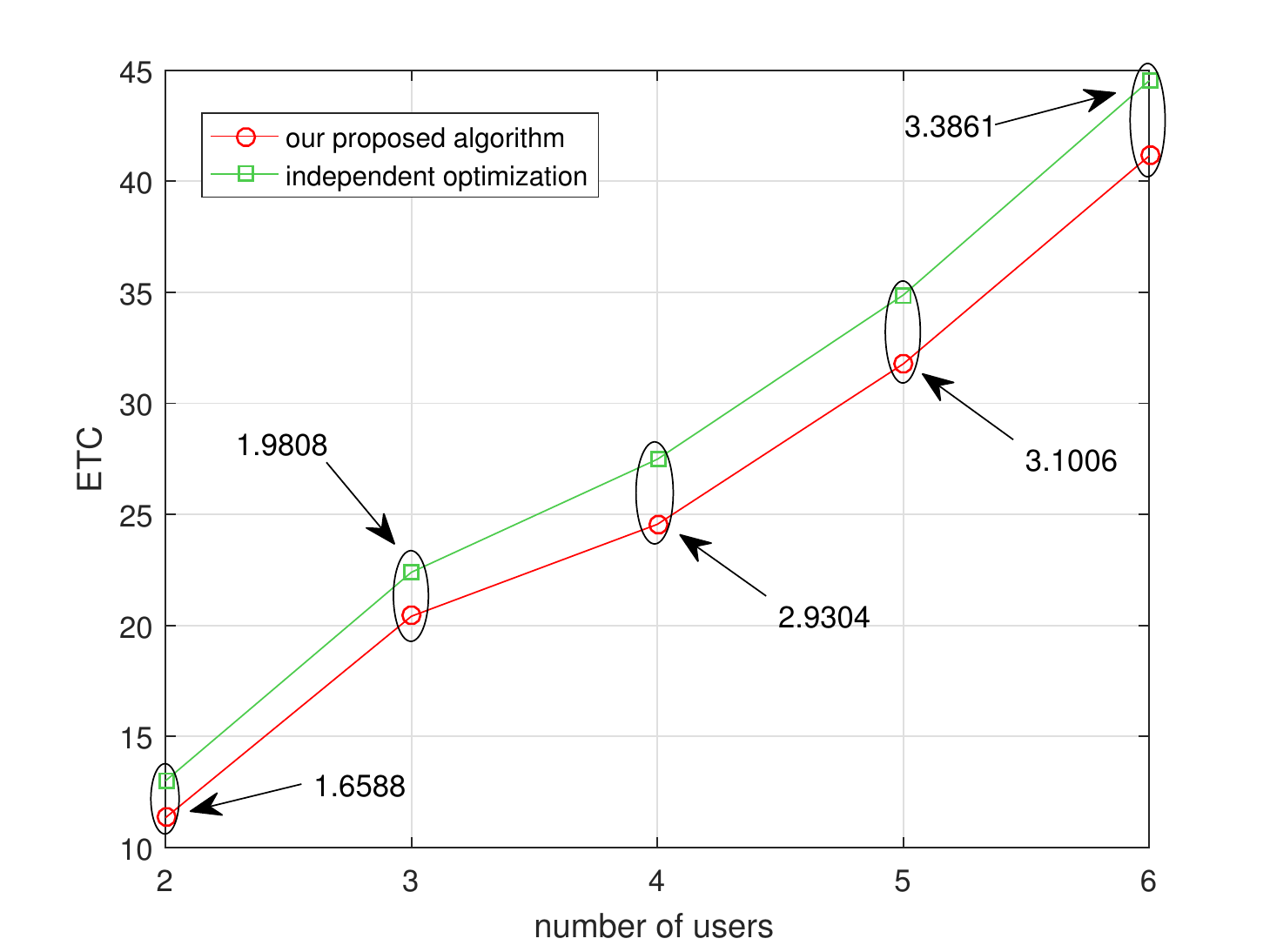}
\vspace{-0.1cm}
 \caption{ Total ETC versus number of users. }
\end{centering}
\vspace{-0.1cm}
\end{figure}

In Fig. 10, we further study the impact of different task dependency model (the call graph) to the system performance. Specifically, given different $\beta_1^T$, Fig. 10 illustrates the variation of the optimal energy and delay costs when the joint task index $k$ changes in the topological call graphs $(3,5,k)$. We observe that with the increase of $k$, $T_1$ becomes larger under a small $\beta_1^T$ (e.g., $\beta_1^T = 0.05$ or $0.3$), while $T_2$ shows an opposite trend. Intuitively, this is because when $k$ is small, e.g., $k=1$ or $k=2$, the inter-user task dependency becomes very stringent, such that WD1 needs to quickly finish all its 3 tasks to meet the finish time of the first $k$ tasks of WD2. At the meantime, WD2 only needs to slow down its computation to ``wait" for WD1's computation results for reduced energy consumption. Overall, this leads to a larger $T_1$ and smaller $T_2$ when $k$ increases.  Besides, when $\beta_1^T$ becomes larger, WD1 pays less emphasis on minimizing its computation delay to meet the computation time of the $k$-th task at WD2. In this case, the computations at the two users are practically decoupled and indeed optimized separately. Therefore, the computation delays at both users are insensitive to the variation of call graph topology, i.e., change of $k$.

In Fig. 11, we illustrate the ETC performance when extending the proposed inter-user task dependency model to the multi-user case, where the distance from each WD to the AP follows a uniform distribution between 10 m and 30 m. Each point in the figure is the average performance of 20 independent distance realizations. Based on the topological call graph in two-user case as shown in Fig. 6, we assume that $\{L_{i,3}\}=[50.5 ~45.3 ~86.6]$ (Mcycles) and $\{O_{i,3}, i=0,1,2,3\}=[1400 ~1200 ~1500 ~1300]$ (KByte) for WD3, $\{L_{i,4}\}=[65.5 ~50.3 ~75.6]$ (Mcycles) and $\{O_{i,4}, i=0,1,2,3\}=[1500 ~1400 ~1000 ~1500]$ (KByte) for WD4,  $\{L_{i,5}\}=[55.5 ~42.3 ~90.6]$ (Mcycles) and $\{O_{i,5}, i=0,1,2,3\}=[1600 ~1500 ~1300 ~1700]$ (KByte) for WD5 and $\{L_{i,6}\}=[58.5 ~47.3 ~82.6]$ (Mcycles) and $\{O_{i,6}, i=0,1,2,3\}=[1200 ~1300 ~1600 ~1600]$ (KByte) for WD6.  The input of the 4-th task at WD2 requires the final task outputs from the other WDs. It is observed that the proposed optimal algorithm outperforms the independent optimization scheme. Specifically, the performance improvement of our proposed algorithm becomes larger when the number of users increases, e.g., from 1.6588 to 3.3861 when user number increases from 2 to 6. It is because the task dependency becomes stronger as the number of users increases, which leads to larger performance gain by considering inter-user task dependency in the optimization.


\begin{figure}
\begin{centering}
\includegraphics[scale=0.6]{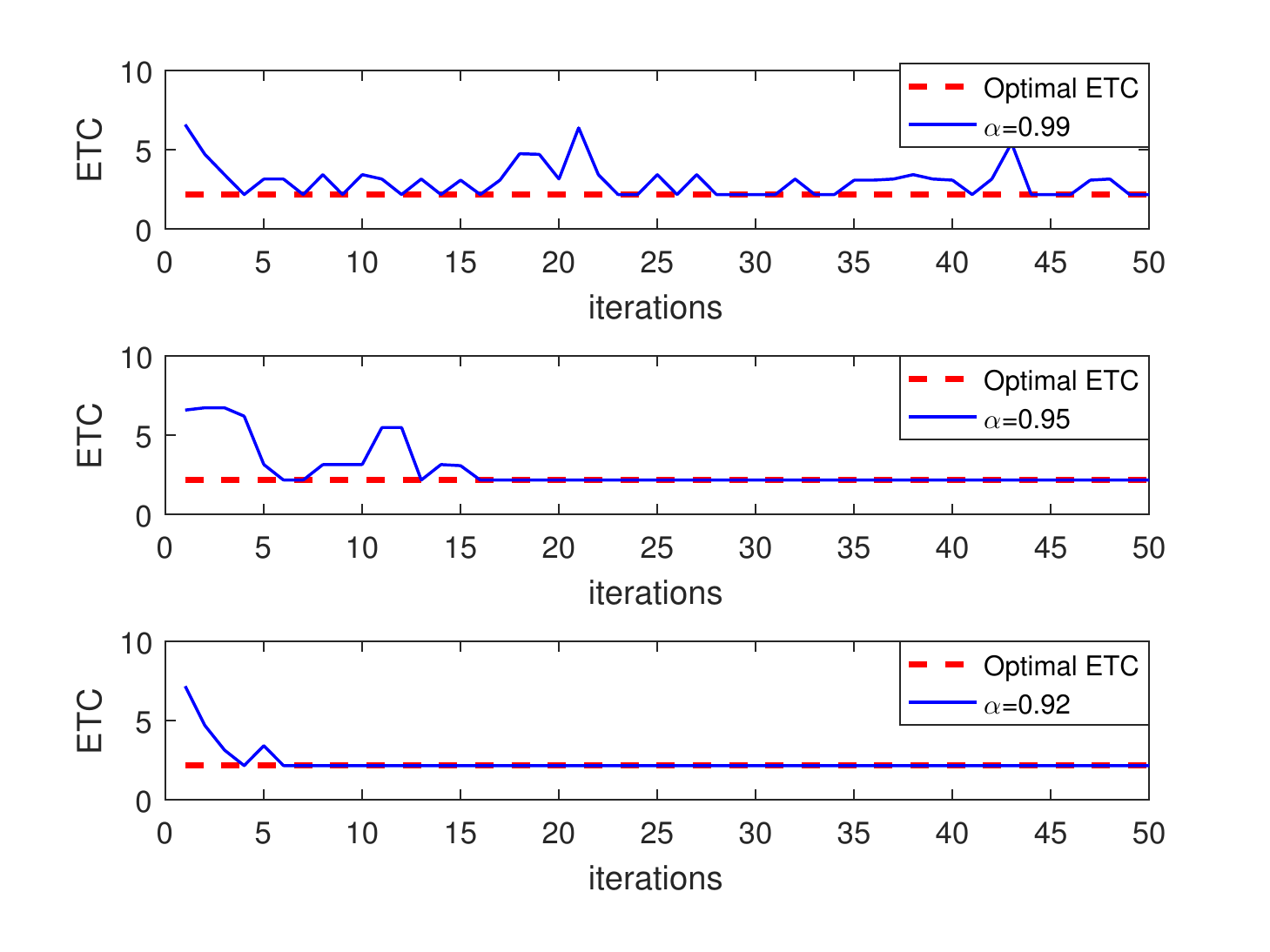}
\vspace{-0.1cm}
 \caption{ Obtained total ETC and number of iterations for different cooling rate $\alpha$. }
\end{centering}
\vspace{-0.1cm}
\end{figure}

\begin{figure}
\begin{centering}
\includegraphics[scale=0.6]{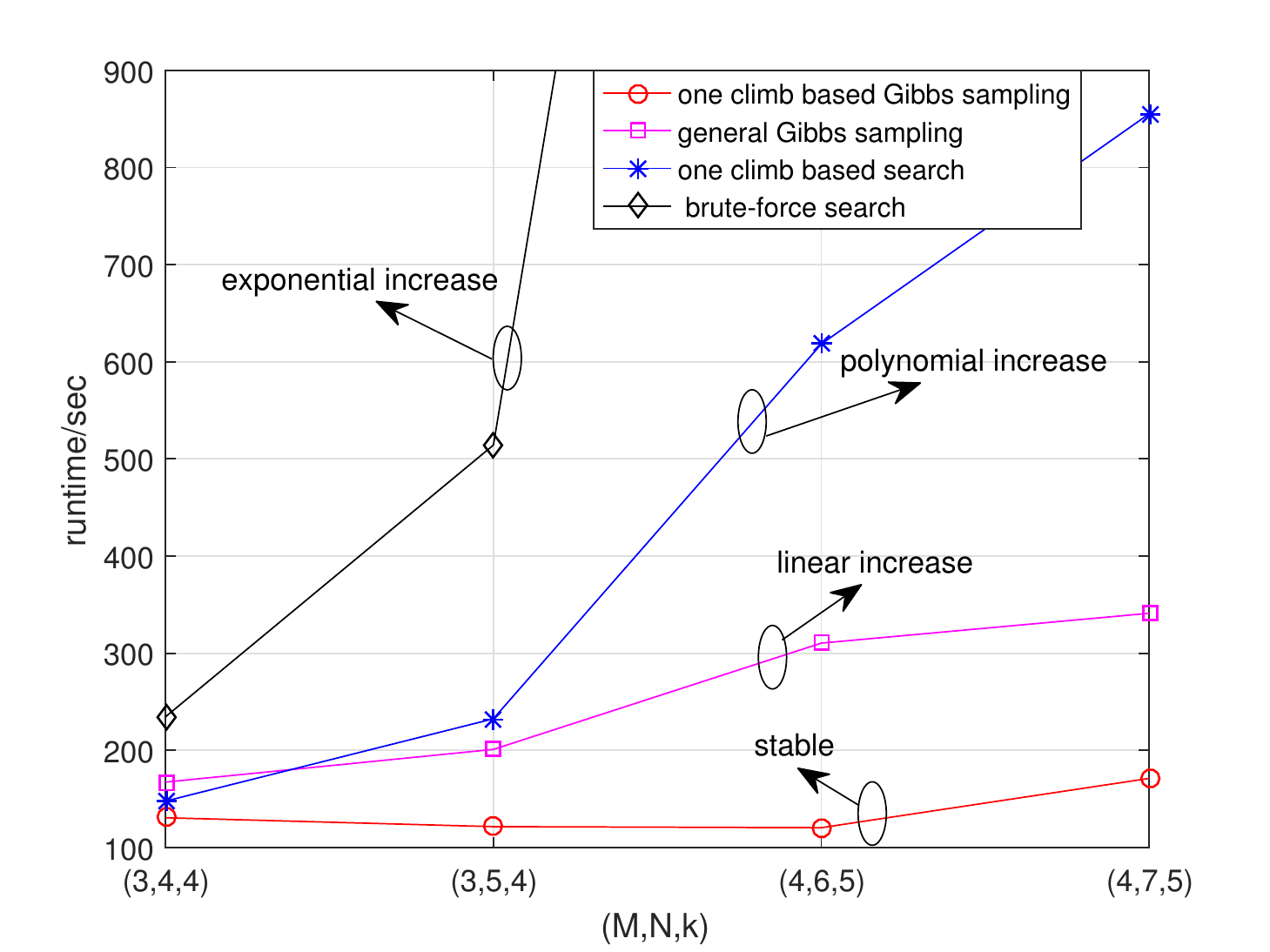}
\vspace{-0.1cm}
 \caption{ Runtime versus different topological call graphs $(M,N,k)$ in four algorithms. }
\end{centering}
\vspace{-0.1cm}
\end{figure}
\subsection{Complexity of the Proposed Gibbs Sampling Algorithm}

In Fig. 12, we plot the obtained total ETC versus the number of iterations, where the initial $T(1)=1$, $d_1=d_2=10$, $\beta_1^T=0$ and $\beta_2^T=0.5$.
It can be seen that the algorithm converges faster with a lower cooling rate. This is because when the temperature $T$ is low, the offloading decision that yields a smaller objective is more likely to be picked, leading to lower fluctuation.

Moreover, in Fig. 13, we compare the complexity among the four algorithms under different topological call graphs $(M,N,k)$, where $d_1=d_2=10$, $\beta_1^T=0$ and $\beta_2^T=0.5$.
The general Gibbs sampling algorithm is considered as a benchmark, where the sampling set is generated without one-climb policy constraint in each iteration. We observe that the proposed one-climb policy based Gibbs sampling algorithm achieves lower runtime compared with the general Gibbs sampling, one-climb policy based search and the brute-force schemes, which indicates the advantages of applying one-climb policy and Gibbs sampling method. Specifically, as the call graph is extended, the brute-force search scheme shows an exponential complexity growth, while the one climb based search method solves the problem in polynomial time. As for the general Gibbs sampling scheme, the runtime is increasing linearly as the size of call graph increases. However, the one climb based Gibbs sampling algorithm is insensitive to the size of call graph and can achieve around $46.68\%$ lower average runtime than the general Gibbs sampling method in Fig. 13.

\section{Conclusions and Future Works}

This paper has studied the impact of inter-user task dependency on the task offloading decisions and resource allocation in a two-user MEC network. We proposed efficient algorithms to optimize the resource allocation and task offloading decisions, with the goal of minimizing the weighted sum of the WDs' energy consumption and task execution time. Besides, we proved that the optimal offloading decisions satisfy an one-climb policy, based on which a reduced-complexity Gibbs sampling algorithm was proposed to obtain the optimal offloading decisions. Simulation results demonstrated that the proposed method can achieve significant performance gain compared to the benchmarks, which indicated the advantage of considering inter-user task dependency. Meanwhile, the complexity of the proposed Gibbs sampling algorithm was low and insensitive to the call graph size.

Finally, we conclude the paper with some interesting future directions. First, we assumed in this paper that each WD is allocated with an orthogonal channel and the CPU frequency of the edge server is fixed. The consideration of both bandwidth and computing resources competitions is needed when we extend our work to a large-size network. In addition, it is interesting to consider more realistic knowledge of channel conditions, where an online optimization algorithm needs to be derived. Besides, there are many other task dependency models and for other more complex models, we can further study them in our future works. Moreover, although the proposed one-climb policy based Gibbs sampling algorithm greatly reduces the computational time compared with the traditional Gibbs sampling algorithm, it may still take a large number of iterations to solve the combinatorial optimization problem. Once the channel conditions change, we need to re-solve the problem. One possible way to address this challenge is to explore the recent development of artificial intelligent algorithms. For example, we can apply the deep reinforcement learning technique to quickly find a mapping between the time-varying channel gains and optimal offloading decisions.

\appendices

\section{Proof of Proposition 3.1} \label{appendicesA}

For the WD1, the derivative of $L$ of \eqref{lagrangian} with respect to $\tau_{i,1}^{l}$ can be expressed as
\begin{align}\nonumber
\frac{\partial L}{\partial\tau_{i,1}^{l}}=\beta_{1}^{T}-\frac{2\kappa\beta_{1}^{E}(L_{i,1})^{3}}{(\tau_{i,1}^{l})^{3}}+\lambda,
\end{align}
 where $\frac{\partial L}{\partial\tau_{i,1}^{l}}$ is a monotonously increasing function with $\tau_{i,1}^{l}\in[\frac{L_{i,1}}{f_{peak}},+\infty)$. Thus, if \\$\frac{\partial L}{\partial\tau_{i,1}^{l}}|_{\tau_{i,1}^{l}=\frac{L_{i,1}}{f_{peak}}}>0$, we have $f_{i,1}^{*}=f_{peak}$. Otherwise,
we have
\begin{align}\nonumber
\tau_{i,1}^{l}=L_{i,1}\sqrt[3]{\frac{2\kappa\beta_{1}^{E}}{\beta_{1}^{T}+\lambda}}\Rightarrow f_{i,1}^{*}=\frac{L_{i,1}}{\tau_{i,1}^{l}}=\sqrt[3]{\frac{\beta_{1}^{T}+\lambda^{*}}{2\kappa\beta_{1}^{E}}}.
\end{align}
Thus,
\begin{align}\nonumber
f_{i,1}^{*}=\min\left\{\sqrt[3]{\frac{\beta_{1}^{T}+\lambda^{*}}{2\kappa\beta_{1}^{E}}},f_{peak}\right\}.
\end{align}
As for the WD2, the proof is similar as that in the WD1 and we omit the details here.

\section{Proof of Proposition 3.2} \label{appendicesB}

In the following, we show the case when $i\leq M$ in the WD1. The proof for the other cases is similar and we omit the details here.

The derivative of $L$ of \eqref{lagrangian} with respect to $\tau_{i,1}^{u}$ is expressed as
\begin{footnotesize}
\begin{align}\nonumber
\frac{\partial L}{\partial\tau_{i,1}^{u}}&=\beta_{1}^{T}+\beta_{1}^{E}\left[\frac{1}{h_{i,1}}f(\frac{O_{i-1,1}}{\tau_{i,1}^{u}})+\frac{\tau_{i,1}^{u}}{h_{i,1}}f'(\frac{O_{i-1,1}}{\tau_{i,1}^{u}})\right]+\lambda\\
&=\beta_{1}^{T}+\beta_{1}^{E}\left[\frac{\sigma^{2}}{h_{i,1}}2^{\frac{O_{i-1,1}}{W\tau_{i,1}^{u}}}(1-\frac{O_{i-1,1}}{W\tau_{i,1}^{u}}\ln2)-\frac{\sigma^{2}}{h_{i,1}}\right]+\lambda.\nonumber
\end{align}
\end{footnotesize}
Next, we can further have the second-order derivative of \eqref{lagrangian} with respect to $\tau_{i,1}^{u}$ as
\begin{align}\nonumber
\frac{\partial^{2} L}{\partial(\tau_{i,1}^{u})^{2}}=\beta_{1}^{E}\frac{\sigma^{2}}{h_{i,1}}\frac{(O_{i-1,1})^{2}}{W^{2}(\tau_{i,1}^{u})^{3}}2^{\frac{O_{i-1,1}}{W\tau_{i,1}^{u}}}(\ln2)^{2}>0,
\end{align}
which indicates that $\frac{\partial L}{\partial\tau_{i,1}^{u}}$ is a monotonously increasing function with \\$\tau_{i,1}^{u}\in[\frac{O_{i-1,1}}{W\log_{2}(1+\frac{P_{peak}h_{i,1}}{\sigma^{2}})},+\infty)$. Let $g=\frac{\partial L}{\partial\tau_{i,1}^{u}}|_{\tau_{i,1}^{u}=\frac{O_{i-1,1}}{W\log_{2}(1+\frac{P_{peak}h_{i,1}}{\sigma^{2}})}}$, we have $\frac{\partial L}{\partial\tau_{i,1}^{u}}\in[g,\beta_{1}^{T}+\lambda]$.

If $g>0$, i.e., $h_{i,1}<\frac{\sigma^{2}}{P_{peak}}[\frac{A_{1}}{-\mathcal{W}(-A_{1}e^{-A_{1}})}-1]$, $L$ is a monotonously increasing function with respect to $\tau_{i,1}^{u}$. Thus, we have $(\tau_{i,1}^{u})^{*}=\frac{O_{i-1,1}}{W\log_{2}(1+\frac{P_{peak}h_{i,1}}{\sigma^{2}})}$, which means that the optimal transmit power of the WD1 in this case is $p_{i,1}^{*}=P_{peak}$. Otherwise, by equating $\frac{\partial L}{\partial\tau_{i,1}^{u}}=0$, we have $p_{i,1}^{*}=\frac{\sigma^{2}}{h_{i,1}}\left[\frac{B_{1}}{\mathcal{W}\left(B_{1}e^{-1}\right)}-1\right]$.

\begin{footnotesize}
\bibliographystyle{IEEEtran}

\begin{thebibliography}{10}
\providecommand{\url}[1]{#1}
\csname url@samestyle\endcsname
\providecommand{\newblock}{\relax}
\providecommand{\bibinfo}[2]{#2}
\providecommand{\BIBentrySTDinterwordspacing}{\spaceskip=0pt\relax}
\providecommand{\BIBentryALTinterwordstretchfactor}{4}
\providecommand{\BIBentryALTinterwordspacing}{\spaceskip=\fontdimen2\font plus
\BIBentryALTinterwordstretchfactor\fontdimen3\font minus
  \fontdimen4\font\relax}
\providecommand{\BIBforeignlanguage}[2]{{%
\expandafter\ifx\csname l@#1\endcsname\relax
\typeout{** WARNING: IEEEtran.bst: No hyphenation pattern has been}%
\typeout{** loaded for the language `#1'. Using the pattern for}%
\typeout{** the default language instead.}%
\else
\language=\csname l@#1\endcsname
\fi
#2}}
\providecommand{\BIBdecl}{\relax}
\BIBdecl

\bibitem{myglobecom}
J.~Yan, S.~Bi, and Y.~J. Zhang, ``Optimal offloading and resource allocation in
  mobile-edge computing with inter-user task dependency,'' \emph{{accepted by}
  IEEE GLOBECOM}, Dec. 2018.

\bibitem{MECsurvey1}
Y.~Mao, C.~You, J.~Zhang, K.~Huang, and K.~B. Letaief, ``A survey on mobile
  edge computing: The communication perspective,'' \emph{IEEE Commun. Surveys
  Tuts.}, vol.~19, no.~4, pp. 2322--2358, Fourthquarter 2017.

\bibitem{MECsurvey2}
W.~Shi, J.~Cao, Q.~Zhang, Y.~Li, and L.~Xu, ``Edge computing: Vision and
  challenges,'' \emph{IEEE Internet Things J.}, vol.~3, no.~5, pp. 637--646,
  Oct. 2016.

\bibitem{bi_DRL}
L.~{Huang}, S.~{Bi}, and Y.~J. {Zhang}, ``Deep reinforcement learning for
  online computation offloading in wireless powered mobile-edge computing
  networks,'' \emph{{IEEE} Trans. Mobile Comput.}, pp. 1--1, 2019.

\bibitem{MEC3}
S.~Bi and Y.~J. Zhang, ``Computation rate maximization for wireless powered
  mobile-edge computing with binary computation offloading,'' \emph{{IEEE}
  Trans. Wireless Commun.}, vol.~17, no.~6, pp. 4177--4190, Jun. 2018.

\bibitem{xu}
F.~Wang, J.~Xu, X.~Wang, and S.~Cui, ``Joint offloading and computing
  optimization in wireless powered mobile-edge computing systems,''
  \emph{{IEEE} Trans. Wireless Commun.}, vol.~17, no.~3, pp. 1784--1797, Mar.
  2018.

\bibitem{MEC2}
C.~You, K.~Huang, and H.~Chae, ``Energy efficient mobile cloud computing
  powered by wireless energy transfer,'' \emph{{IEEE} J. Sel. Areas Commun.},
  vol.~34, no.~5, pp. 1757--1771, May 2016.

\bibitem{MEC5}
W.~Zhang, Y.~Wen, K.~Guan, D.~Kilper, H.~Luo, and D.~O. Wu, ``Energy-optimal
  mobile cloud computing under stochastic wireless channel,'' \emph{{IEEE}
  Trans. Wireless Commun.}, vol.~12, no.~9, pp. 4569--4581, Sept. 2013.

\bibitem{MEC7}
M.~H. Chen, B.~Liang, and M.~Dong, ``Joint offloading decision and resource
  allocation for multi-user multi-task mobile cloud,'' in \emph{Proc. IEEE
  ICC}, May 2016.

\bibitem{MEC4}
T.~Q. Dinh, J.~Tang, Q.~D. La, and T.~Q.~S. Quek, ``Offloading in mobile edge
  computing: Task allocation and computational frequency scaling,''
  \emph{{IEEE} Trans. Commun.}, vol.~65, no.~8, pp. 3571--3584, Aug. 2017.

\bibitem{HongXing}
H.~Xing, L.~Liu, J.~Xu, and A.~Nallanathan, ``Joint task assignment and
  wireless resource allocation for cooperative mobile-edge computing,'' in
  \emph{Proc. IEEE ICC}, May 2018.

\bibitem{cs_taskgraph}
Y.-K. Kwok and I.~Ahmad, ``Dynamic critical-path scheduling: an effective
  technique for allocating task graphs to multiprocessors,'' \emph{{IEEE}
  Trans. Parallel Distrib. Syst.}, vol.~7, no.~5, pp. 506--521, May 1996.

\bibitem{add1}
M.~A. {Rodriguez} and R.~{Buyya}, ``Deadline based resource provisioningand
  scheduling algorithm for scientific workflows on clouds,'' \emph{IEEE
  Transactions on Cloud Computing}, vol.~2, no.~2, pp. 222--235, Apr. 2014.

\bibitem{add2}
Z.~{Wu}, Z.~{Ni}, L.~{Gu}, and X.~{Liu}, ``A revised discrete particle swarm
  optimization for cloud workflow scheduling,'' in \emph{2010 International
  Conference on Computational Intelligence and Security}, Dec 2010, pp.
  184--188.

\bibitem{add3}
S.~{Pandey}, L.~{Wu}, S.~M. {Guru}, and R.~{Buyya}, ``A particle swarm
  optimization-based heuristic for scheduling workflow applications in cloud
  computing environments,'' in \emph{2010 24th IEEE International Conference on
  Advanced Information Networking and Applications}, April 2010, pp. 400--407.

\bibitem{single1}
S.~B. P.~D.~Lorenzo and S.~Sardellitti, ``Joint optimization of radio resources
  and code partitioning in mobile edge computing,'' \emph{\normalfont{submitted
  for publication, available on-line at http://arxiv.org/abs/1307.3835v3}}.

\bibitem{single4}
W.~Zhang, Y.~Wen, and D.~O. Wu, ``Collaborative task execution in mobile cloud
  computing under a stochastic wireless channel,'' \emph{{IEEE} Trans. Wireless
  Commun.}, vol.~14, no.~1, pp. 81--93, Jan. 2015.

\bibitem{single5}
W.~Zhang and Y.~Wen, ``Energy-efficient task execution for application as a
  general topology in mobile cloud computing,'' \emph{{to appear in} IEEE
  Transactions on Cloud Computing}.

\bibitem{single2}
M.~Jia, J.~Cao, and L.~Yang, ``Heuristic offloading of concurrent tasks for
  computation-intensive applications in mobile cloud computing,'' in
  \emph{Proc. IEEE INFOCOM WKSHPS}, Apr. 2014.

\bibitem{multi1}
S.~Guo, B.~Xiao, Y.~Yang, and Y.~Yang, ``Energy-efficient dynamic offloading
  and resource scheduling in mobile cloud computing,'' in \emph{Proc. IEEE
  INFOCOM}, Apr. 2016.

\bibitem{DL2}
R.~{Viswanathan} and P.~K. {Varshney}, ``Distributed detection with multiple
  sensors part i. fundamentals,'' \emph{Proc. {IEEE}}, vol.~85, no.~1, pp.
  54--63, Jan 1997.

\bibitem{DL}
J.~B. {Predd}, S.~B. {Kulkarni}, and H.~V. {Poor}, ``Distributed learning in
  wireless sensor networks,'' \emph{IEEE Signal Processing Magazine}, vol.~23,
  no.~4, pp. 56--69, July 2006.

\bibitem{sensor1}
I.~A. et~al., ``Wireless sensor networks: A survey,'' \emph{Computer Networks,
  Elsevier Science}, vol.~38, no.~4, pp. 393--422, 2002.

\bibitem{sensor2}
Y.~.~E. {Wang}, X.~{Lin}, A.~{Adhikary}, A.~{Grovlen}, Y.~{Sui},
  Y.~{Blankenship}, J.~{Bergman}, and H.~S. {Razaghi}, ``A primer on 3gpp
  narrowband internet of things,'' \emph{IEEE Communications Magazine},
  vol.~55, no.~3, pp. 117--123, March 2017.

\bibitem{convex}
S.~Boyd and L.~Vandenberghe, \emph{\normalfont{Convex Optimization}}.\hskip 1em
  plus 0.5em minus 0.4em\relax Cambidge University Press, 2004.

\bibitem{GS1}
S.~Geman and D.~Geman, ``Stochastic relaxation, gibbs distributions, and the
  bayesian restoration of images,'' \emph{{IEEE} Trans. Pattern Anal. Mach.
  Intell.}, vol. PAMI-6, no.~6, pp. 721--741, Nov 1984.

\bibitem{GS2}
L.~P. Qian, Y.~J.~A. Zhang, and M.~Chiang, ``Distributed nonconvex power
  control using gibbs sampling,'' \emph{{IEEE} Trans. Commun.}, vol.~60,
  no.~12, pp. 3886--3898, Dec 2012.

\bibitem{GS3}
D.~Bertsimas and J.~Tsitsiklis, ``Simulated annealing,'' \emph{Statistical
  Science}, vol.~8, no.~1, pp. 10--15, 1993.

\bibitem{parameter1}
A.~P. Miettinen and J.~K. Nurminen, ``Energy efficiency of mobile clients in
  cloud computing,'' in \emph{Proc. 2nd USENIX Conf. Hot Topics Cloud Comput.},
  Jun. 2010, pp. 4--11.

\bibitem{parameter2}
Y.~Xiao, P.~Savolainen, A.~Karppanen, M.~Siekkinen, and A.~Yla-Jaaski,
  ``Practical power modeling of data transmission over 802.11g for wireless
  applications,'' in \emph{Proc. 1st Int. Conf. EnergyEfficient Comput. Netw.},
  2010, pp. 75--84.

\end{thebibliography}

\end{footnotesize}

\end{document}